\documentclass[a4paper,11pt]{article}
\usepackage{jcappub} % for details on the use of the package, please
\usepackage[T1]{fontenc} % if needed

\title{Decaying vacuum energy, matter creation and cosmic acceleration}

\author{Lokesh Chander}
\author{and C.P. Singh}
\affiliation{Department of Applied Mathematics, Delhi Technological University, Delhi - 110042, India}

\emailAdd{lokeshchander\_23phdam04@dtu.ac.in}
\emailAdd{cpsingh@dce.ac.in}
\abstract{We discuss an interacting dark sector model featuring decaying vacuum energy and dark matter empowered by gravitationally induced matter creation. Motivated by quantum field theoretic considerations of vacuum decay and adiabatic particle production, we analyse both the background dynamics and the growth rate of perturbations. The model is confronted with diverse datasets, including Cosmic Chronometers, Pantheon Type Ia Supernovae, Baryon Acoustic Oscillations, Cosmic Microwave Background distance priors, weighted linear growth rate measurements and an $H_0$ prior, with parameter estimation performed via Markov Chain Monte Carlo (MCMC) methods. Model comparison is carried out using the Akaike and Deviance Information Criterion. Our results show a consistent transition from a decelerated to an accelerated expansion phase, with present Hubble parameter estimates lying between the Planck and SH0ES values, thereby easing the Hubble tension. The structure growth parameter $S_8$ is also compatible with Planck 2018 and recent weak lensing surveys. A thermodynamic analysis confirms consistency with the generalized second law, and including Casimir contributions provides further insights into the model's dynamics. Overall, the proposed model effectively captures the Universe's evolution at both theoretical and observational levels.}

\begin{document}
\maketitle
\flushbottom

\section{Introduction}
\label{sec1}
The first observational evidence for the accelerated expansion of the Universe was confirmed through Type Ia supernovae (SNIa), rightly called `standard candles', as these astronomical objects have an intrinsic luminosity ~\cite{Riess98,Perlmutter99}. This discovery significantly changed our understanding of the Universe and caused a resurgence in modern cosmology. Subsequent datasets: Baryon Acoustic Oscillations (BAO) ~\cite{Eisenstein05a}, the Hubble parameter measurements $H(z)$ ~\cite{Farooq13}, and Cosmic Microwave Background (CMB) anisotropies ~\cite{Larson11} have further validated this phenomenon. According to `General Relativity' (GR), this accelerated expansion is driven by a new energy density with negative pressure, referred to as `dark energy'(DE), which makes up around 70\% of the Universe's total energy density. Accompanying this is `dark matter'(DM), whose nature is not entirely understood and together, both the dark components, collectively called the `dark sector', account for nearly 95\% of the Universe's content ~\cite{Sahni04}. Understanding the origin of the dark sector remains one of the major unsolved issues in modern cosmology. The naturally arising candidate for dark energy within the framework of GR is the cosmological constant, denoted by $\Lambda$, or equivalently, the vacuum energy density associated with it, $\rho_{\Lambda}=\Lambda/8\pi G$ ~\cite{ORaif18}. It has a negative pressure, making it the repulsive driving force that causes accelerated expansion in late-time, under the condition that $\Lambda$ has a value greater than zero ~\cite{Sahni00}.\\
\indent However, the $\Lambda$-based concordance model faces theoretical challenges, most notably the cosmological constant problem and cosmic coincidence ~\cite{Krauss95,Carroll01,Schutz02,Weinberg89,Velten14}, which remain central issues in cosmology. These unresolved problems have motivated alternative approaches, such as dynamic dark energy models, where $\Lambda$ evolves as a function of time, and interacting models, where dark energy interacts with other components in the cosmic fluid. In decaying vacuum energy models, $\Lambda$ is not a constant; instead, it evolves over time and is thus a dynamic component $\Lambda(t)$. This approach addresses the fine-tuning problem by linking the present-day small value of $\Lambda$ to its evolutionary trajectory during the Universe's history. Early studies ~\cite{Berto86,Freese87,Lima96,Overduin98,Shap00,Shap02,Alcaniz03,Opher04,Bauer05,Carn05,Alcaniz05,Sola05,Barrow06,Monte07,Basil09a,Shap09}, often motivated by phenomenological considerations, laid the theoretical foundations and proposed certain functional forms of $\Lambda(t)$. More recently, many authors~\cite{Tong11,Sola11,Lima13,Szy15,Lima15,Zil18,Singh21,Singh24} have sought to ground these models in quantum field theory (QFT). Such QFT-motivated decaying vacuum models are often referred to as Running Vacuum Models (RVMs). Within this framework, $\Lambda(t)$ can also be interpreted as a remnant of the inflationary phase~\cite{Sola08}. Notably, recent developments have demonstrated that inflation itself can be generated from vacuum decay into matter in the context of RVMs, derived from QFT in curved spacetime~\cite{Sola25}, thus offering a unified theoretical framework that connects the early and late accelerated expansion of the Universe.\\
\indent Besides the dynamical $\Lambda(t)$ approach, recently, a new alternative to address late-time cosmic acceleration has gained traction. It is the theory of `adiabatic' matter creation, which is gravitationally induced and a non-equilibrium thermodynamic process. In 1939, Schr\"{o}dinger ~\cite{Schro39} conceptualised matter creation in the microscopic realm, which was further developed by Parker in the late 1960s~\cite{Park68,Park69}. Since Einstein's field equations are the background equations to understand the evolution of the Universe, it was Prigogine and his collaborators ~\cite{Prig86,Prig88,Prig89} who proposed a thermodynamic framework for open systems, allowing the creation of matter via transfer of energy from the gravitational field. In this case, by assuming that DM is being produced due to a time-varying gravitational field, it is plausible for a late-time acceleration to happen in a Universe consisting of only pressureless fluids, like cold dark matter and baryons. The research work carried out by many authors ~\cite{Calvao92,Lima92,Lima96a,Zim96,Johri96,Alcaniz99,Zim01,Yaun07,Steig09,Singh11,Singh12,Lima12,Lima14,Ramos14,Pan18,Singh20,Bhar24} suggests that matter creation can mimic the effects of dark energy, producing negative pressure and driving accelerated expansion. These models, rooted in thermodynamics, posit that new matter gets created as the Universe expands. With the energy conservation law, the creation of new particles may come at the expense of another component in the cosmic fluid; with the most likely canditate being the vacuum. The idea of vacuum decay leading to particle production was explored by Rubakov in the seminal work~\cite{Rub84}. In a scenario with matter creation, a decaying vacuum term $\Lambda(t)$ can lead to a naturally arising coupling between the DE and DM sector. The decay of $\Lambda(t)$ is thus tied to the creation of matter via the fluid equation. While the exact mechanism of this phenomenon has not been deduced, the implication is clear that DE and DM are interacting entities.\\
\indent Another viable alternative is provided by the exotic fluid called Chaplygin gas (CG), which behaves like DM at early redshifts and like DE at later times. Many models based on various forms of the CG have been studied in the literature, and this approach has been promising since it is consistent with observational data~\cite{Guo07,Zheng22,Salah20}. Moreover, there are interacting dark energy (IDE) models in which interaction between the DE and other components in the cosmic fluid is assumed, and these models have been extensively explored in the literature. Along with the class of CG models, these models can solve the coincidence problem by allowing dark energy and dark matter to evolve over time. The gradual convergence of the two densities helps to resolve the issue of their seemingly fine-tuned values at present~\cite{Chim03,Amen07,Chow13,Zhang09}.\\
\indent The present study investigates a combined model incorporating decaying vacuum energy and gravitational matter creation, offering a unified framework to address the aforementioned challenges. Unlike some interacting dark energy models that require an explicitly defined coupling term, this approach inherently links the decay of $\Lambda(t)$ to the rate of matter creation, denoted by $\Gamma$, avoiding the need for arbitrary assumptions about the form of coupling. To test the viability of the proposed model, we undertake a rigorous theoretical and observational analysis. With the help of Bayesian Markov Chain Monte Carlo (MCMC) methods, the model parameters are constrained with multiple datasets, including Baryon Acoustic Oscillations (BAO), Cosmic Microwave Background (CMB) distance priors, Pantheon Supernovae Type Ia (SNIa), H(z) measurements from Cosmic Chronometers (CC), and weighted linear growth rate measurements $f(z)\sigma_8(z)$. We also incorporate a prior on the Hubble constant $H_0$ from the R22 local measurement~\cite{Riess22} to examine the Hubble tension. Further, we employ statistical criteria: the reduced chi-squared test and two model selection criteria, AIC and DIC, to test the model against a reference. This allows for a robust comparison with the standard $\Lambda$CDM model, highlighting the potential advantages and limitations of the proposed framework. The thermodynamic analysis has also been carried out to test the validity of the generalized second law of thermodynamics for the model.\\
\indent The paper is organised as follows: Section~\ref{sec2} establishes the theoretical framework of our cosmological model, which combines decaying vacuum energy with matter creation, and outlines the corresponding modifications to the energy–momentum tensor and the Friedmann equations. In Section~\ref{sec3}, we derive analytic solutions to the field equations and examine key cosmological parameters, namely the Hubble parameter $H(z)$, the deceleration parameter $q(z)$, the transition redshift $z_{\text{tr}}$, and the effective equation of state parameter $w_{\text{eff}}(z)$. Section~\ref{sec4} is devoted to the study of cosmic perturbations, with emphasis on matter fluctuations and their role in structure formation. Section~\ref{sec5} describes the observational datasets and the methodology employed to constrain the model, while Section~\ref{sec6} presents the corresponding results and their implications. In Section~\ref{sec7}, we assess the model’s statistical performance relative to $\Lambda$CDM using the reduced chi-squared test and two information criteria. Section~\ref{sec8} turns to the thermodynamic analysis: we derive an expression for the dark matter temperature, evaluate the evolution of the total entropy of the universe, and test the validity of the generalised second law of thermodynamics, explicitly incorporating the Casimir effect. Finally, Section~\ref{sec9} summarises our conclusions and outlines directions for future research.

\section{Decaying Vacuum Model with Matter Creation}
\label{sec2}
Let us begin with a theoretical framework characterized by a spatially isotropic and homogeneous flat Friedmann-Lema\^{i}tre-Robertson-Walker line element:
\begin{equation}
	ds^2 = -dt^2 + a^2(t) \left[ dr^2 + r^2 \left( d\theta^2 + \sin^2\theta \, d\phi^2 \right) \right],
	\label{eq:FLRW_metric}
\end{equation}
where $(t, r, \theta, \phi)$ represent spherical coordinates with $t$ denoting the cosmic time and $a(t)$ is the scale factor. The convention with relativistic units $c=1$ is assumed in this study.\\
\indent We introduce the basic thermodynamic concept of adiabatic matter creation in a cosmological context as proposed by Prigogine and his collaborators~\cite{Prig86,Prig88,Prig89}. We consider an $N$ particle system under adiabatic conditions, which means that the total particle number changes and is not conserved. Mathematically, we express it as $N^\mu_{\ ;\mu} \neq 0 $. Due to this, the energy conservation equation gets modified. The particle flux vector, which measures how the particle number changes, is taken as $N^\mu = n u^\mu$, where $n = N/V$ gives the number of particles per unit of time (particle number density), and $u^\mu$ is the fluid's four-velocity.\\
\noindent The non-conservation of the particle flux, i.e., $\nabla_\mu n^\mu=\psi$, can be written as
\begin{equation}
	\dot{n}+3Hn=\psi = n\Gamma,
	\label{balance}
\end{equation}
where $\dot{}$ denotes differentiation with respect to cosmic time, $H=\dot{a}/a$ is the Hubble parameter, and $\Gamma$ quantifies the rate of change of the particle number in a physical volume $V$ consisting of $N$ particles. Although the physical nature of $\Gamma$ is unknown, we have one constraint over $\Gamma$, i.e., $\Gamma\ge 0$. For the generalized second law of thermodynamics to be valid, this must be true. Eq.~\eqref{balance} gives the evolution of the particle number density. As the total number of particles per comoving volume $a^3$ is $N=a^3n$, therefore, the above Eq.~\eqref{balance} can be written as $\dot{N}/N=\Gamma$. We define the entropy flux vector as $s^\mu = n\sigma u^\mu$, where $\sigma$ is the entropy per particle. The entropy must satisfy the second law of thermodynamics, $s^\mu_{;\;\mu} \geq 0$. Also, we consider matter creation to be an adiabatic process, i.e., we have a system in which $\sigma$ remains constant.\\
\indent As we are assuming the total particle number $N$ to be variable, we have a modification of the first law of thermodynamics:
\begin{equation}
	d(\rho V) + p \, dV - \left(\frac{h}{n}\right) d(nV) = 0, \label{eq:first_law_modified}
\end{equation}
where $\rho$ is the energy density and $p$ is the equilibrium pressure of the cosmic matter fluid (dark matter and baryonic matter), and $h = (\rho + p)$ is the enthalpy per unit volume. We observe there to be an extra contribution in  \eqref{eq:first_law_modified}, and we interpret this as a non-thermal pressure arising due to the particle production process. We call this the `creation pressure' $p_c$, which is given in terms of the change in particle number with respect to volume by
\begin{equation}
	p_c = -\frac{h}{n} \frac{d(nV)}{dV}.
	\label{eq:creation_pres}
\end{equation}
The negative sign indicates that $p_c$ has a similar nature to the dark energy pressure and may contribute to cosmic expansion. For simplicity's sake, we assume that only dark matter particles are produced since the production of baryons is limited by the local gravity measurement-induced constraints. \\
\indent From \eqref{balance} and \eqref{eq:creation_pres}, along with the modified first law in \eqref{eq:first_law_modified}, we obtain the following form of the creation pressure, derived in the works~\cite{Prig89,Calvao92,Bhar24}:
\begin{equation}
	p_c = -(\rho + p) \frac{\Gamma}{3H}.
	\label{eq:creation_pressure_final}
\end{equation}
Now, let us outline the fundamental equations of a general cosmological scenario. The gravitational field equations in Einstein's General Relativity are:
\begin{equation}
G_{\mu\nu} = \kappa \tilde{T}_{\mu\nu},
	\label{eq:EFE}
\end{equation}
where $G_{\mu\nu}=R_{\mu\nu} - \frac{1}{2} g_{\mu\nu}R$ is the Einstein tensor which describes the spacetime geometry and $\kappa$ is the Einstein gravitational constant which in this case is, $\kappa=8\pi G$. $\tilde{T}_{\mu\nu}\equiv T_{\mu\nu}+ g_{\mu\nu}\rho_\Lambda$ is the total energy-momentum tensor, which accounts for the contribution of both matter creation and a time-dependent vacuum term. In the presence of matter creation, it is customary to associate the creation pressure, $p_c$, with the energy-momentum tensor. As a result, the usual perfect fluid energy-momentum tensor $T_{\mu\nu}$, gets modified to
\begin{equation}
	T_{\mu\nu} = \left( \rho_m + \tilde{p}_m \right) u_\mu u_\nu + g_{\mu\nu}\tilde{p}_m,
	\label{eq:energy_momentum_tensor}
\end{equation}
where $u_\mu$ is the four-velocity of the matter fluid, $\rho_m$ is the matter energy density, and $\tilde{p}_m$ is the total pressure exerted by the fluid, which is the sum of the isotropic pressure $p_m$ and the creation pressure $p_c$. We assume matter to be dust-like, i.e. it has negligible pressure ($p_m=0$). In this paper, we assume that the vacuum behaves like a perfect fluid, with an equation of state given by $\rho_{\Lambda}=-p_{\Lambda}$. Given these considerations, we can write the Friedmann equations as follows:
\begin{equation}
    3 H^2 = 8 \pi G (\rho_m + \rho_\Lambda),
    \label{eq:FE1}
\end{equation}
\begin{equation}
    2\dot{H} + 3 H^2 = -8 \pi G (p_c + p_\Lambda),
    \label{eq:FE2}
\end{equation}
where $H = \dot{a}/a$ gives the expansion rate and is called the Hubble parameter. The conservation of the energy-momentum tensor, $ \nabla^\mu \tilde{T}_{\mu \nu} = 0$ gives
\begin{equation}\label{cons}
	\dot{\rho}_m + \dot{\rho}_\Lambda + 3H \left(\rho_m + \rho_\Lambda + p_c + p_\Lambda \right) = 0.
\end{equation}
Since \(\rho_\Lambda = - p_\Lambda \), the equation \eqref{cons} gets simplified to:
\begin{equation}
	\dot{\rho}_m + 3H \left(\rho_m + p_c \right) = -\dot{\rho}_\Lambda.
	\label{eq:conteq}
\end{equation}
From \eqref{eq:FE1} and \eqref{eq:conteq}, we can write the evolution equation as
\begin{equation}\label{evolution eq}
2\dot{H} + 3H^2 + 8\pi G p_c = 8\pi G \rho_\Lambda.
\end{equation}

\section{Solution of the field equations}
\label{sec3}
In this section, we find the solution of the evolution equation \eqref{evolution eq} in the form of the Hubble parameter $H(z)$, which allows us to constrain the model parameters. Eq.~\eqref{evolution eq} can be solved for $H$ if we know $\rho_{\Lambda}$, and $\Gamma$ which gives us $p_c$ through Eq.~\eqref{eq:creation_pressure_final}. Parametrization of $\Lambda$ has changed over time and some notable forms are:  $\Lambda \propto a^{-2}$ ~\cite{Ozer87}, $\Lambda \propto a^{-m}$ ~\cite{Gasp88} and $\Lambda \propto H^2$ ~\cite{Carv92}. Although there is no unique or universally accepted form for the decaying vacuum energy density (VED), it is often modeled as a function of the scale factor $a$ or the Hubble expansion rate $H$. In particular, a dependence on $H^2$ has been motivated by early works in quantum field theory (QFT) in curved spacetime~\cite{Adler82, Park85} and later formalized within the framework of the renormalization group (RG) approach ~\cite{Basil09b,Sola13,Shap09,Oli14}. The influential work by Shapiro and Sol\`{a}~\cite{Shap02}, further developed by several authors~\cite{Grande11,Bess13,Jaya19,Singh24}, supports the idea of running vacuum models (RVMs) which have a dynamical vacuum component. More recently, a series of theoretical studies have derived this vacuum evolution law from QFT calculations in curved spacetime, strengthening its physical basis~\cite{gv20,Sola22,mp23,SolaRev22}. Motivated by these developments, we adopt the following parametrization for the vacuum energy density:
\begin{equation}
	\rho_\Lambda = \frac{3}{8 \pi G} (c_0 + \nu H^2),
	\label{eq:vac_density}
\end{equation}
where $c_0$ is an additive constant term, and $\nu$ is a dimensionless parameter characterizing the dynamical evolution of vacuum energy, which is consistent with theoretical expectations from RG approaches~\cite{Sola18b}. The parameter $\nu$ is constrained by observational data and is expected to be small, i.e., $|\nu| \ll 1$.
In the limit $\nu \to 0$, the vacuum energy density becomes a constant, i.e. it reduces to the cosmological constant. This form is theoretically motivated due to its relation to the general form of the effective action of QFT in curved spacetime. \\
\indent In the proposed model, the cosmic history is based on the rate of expansion given by $H$ and the evolution of energy density, which can define the matter creation rate $\Gamma$. A simple choice for the creation rate is to take it as a constant, $\Gamma=\Gamma_0$, where matter creation requires $\Gamma_0>0$. Another possibility is to assume a direct proportionality to the Hubble parameter, $\Gamma \propto H$. In this paper, we consider a simple form that scales directly with the Hubble parameter~\cite{Lima96a} as
\begin{equation}\label{rate}
	\Gamma = 3\beta H.
\end{equation}
The parameter $\beta$ is associated with the particle production. It is important to note that the matter creation rate $\Gamma$ drops to zero in the limit $\beta \to 0$, meaning that the standard model with matter conservation (no creation or annihilation) is restored. Thus, its value is expected to be small and positive.\\
\indent Since the proposed framework has a naturally arising interaction due to the vacuum decay, we write the continuity equations for the matter and vacuum components in the following way:
\begin{equation}\label{cont_eq_matter}
  \dot{\rho}_m + 3 H (1-\beta) \rho_m = Q, \quad
  \dot{\rho}_{\Lambda} = -Q.
\end{equation}
Here, $Q$ is the interaction term, and its expression comes directly from the vacuum decay as
\begin{equation}\label{interaction_term}
  Q = - \frac{6 \nu \dot{H} H}{8 \pi G} .
\end{equation}
The sign of $Q$ determines the direction of energy flow. In the proposed scenario, vacuum energy decaying into dark matter is plausible because it aligns with particle production. Moreover, a model where the energy flow is in the opposite direction is generally not viable ~\cite{van2024}, and therefore, we must ensure that $Q$ is positive. We note that $Q \propto -6 \dot{H} H / 8 \pi G = - (\dot{\rho}_m + \dot{\rho}_{\Lambda})$ and $\nu$, in addition to being the vacuum decay parameter, also determines the strength of the interaction. For the vacuum to act as a source, we must have $Q > 0$, and since $\dot{H}$ is negative, the viability condition gives us $\nu > 0$. Additionally, from the above, we interpret matter to have an effective equation of state $w_m = -\beta$ due to the adiabatic matter creation process, which indicates a small deviation in behavior from a dust-like fluid.\\
\indent Using \eqref{eq:vac_density} and \eqref{rate} and substituting in \eqref{evolution eq}, we obtain
\begin{equation}
	\dot{H} + \frac{3}{2} (1 - \beta)(1 - \nu) H^2 = 3H_0^2 (1 - \beta)(\Omega_\Lambda - \nu),
	\label{expr1}
\end{equation}
where $c_0=H^2_0(\Omega_{\Lambda}-\nu)$, $H_0$ is the value of $H$ at present and is called the Hubble constant, and $\Omega_{\Lambda}$ is the present value of the dark energy density parameter. Eq.~\eqref{expr1} is a differential equation in $H$ with respect to cosmic time $t$. Instead of using the variable $t$, it is more convenient to write the above equation in terms of the scale factor $a$ or redshift $z$.  Thus, Eq.~\eqref{expr1} can be rewritten as
\begin{equation}\label{exp2}
	\frac{dH^2}{da} + \frac{3}{a}(1 - \beta)(1 - \nu) H^2 = \frac{3}{a}(1 - \beta)(\Omega_\Lambda - \nu) H_0^2.
\end{equation}
The solution of \eqref{exp2} is given by
\begin{equation}
	H(z) = H_0 \left[\left(1 - \frac{\Omega_\Lambda - \nu}{1 - \nu}\right)(1 + z)^{3(1 - \beta)(1 - \nu)} + \frac{\Omega_\Lambda - \nu}{1 - \nu}\right]^{1/2},
	\label{HubbleParameter}
\end{equation}
where $(1+z)=a_0/a(t)$. Here, $a_0$ represents the current cosmic scale factor value, typically assumed to be $1$. The above expression gives the expansion rate in terms of the four independent model parameters $H_0$, $\Omega_\Lambda$, $\nu$ and $\beta$. Since we have considered only matter and vacuum components to be part of the cosmic fluid, we have $\Omega_m + \Omega_\Lambda = 1$.\\
\indent While similar analytic solutions for $H(z)$, as in Eq.~\eqref{HubbleParameter}, have been presented in the context of running vacuum models (RVMs)~\cite{Shap03b,Esp04}, our formulation introduces an additional physical mechanism in the form of matter creation, characterized by a production rate $\Gamma = 3\beta H$. This modifies the continuity equation for our unified dark sector and alters the overall expansion dynamics. As a result, the proposed model represents a broader class that generalizes the standard RVM framework and is thereby distinguished from earlier models. Hereafter, we denote this framework as $\Gamma$RVM. \\

\noindent From \eqref{HubbleParameter}, the scale factor in terms of $t$ can be found as
\begin{equation}\label{scale}
a=\left[\sqrt{\frac{\tilde{\Omega}_m}{\tilde{\Omega}_{\Lambda}}}\sinh \left(\frac{3}{2}\sqrt{\tilde{\Omega}_{\Lambda}}(1-\beta)(1-\nu)H_0t\right)\right]^{\frac{2}{3(1-\beta)(1-\nu)}},
\end{equation}
where $\tilde{\Omega}_m=1 - \frac{\Omega_\Lambda - \nu}{1 - \nu}$ and $\tilde{\Omega}_{\Lambda}=\frac{\Omega_\Lambda - \nu}{1 - \nu}$ are the effective fractional energy density parameters for matter and dark energy respectively.\\
\indent To discuss the decelerating and accelerating phases of the expansion of the Universe, along with the transition, we study the cosmological parameter known as the deceleration parameter, $q$, which is defined as:
\begin{equation}
	q = -\frac{\ddot{a}}{aH^2} = -\left(1 + \frac{\dot{H}}{H^2}\right).
	\label{eq:deceleration_parameter}
\end{equation}
\\
The equation \eqref{HubbleParameter} is used for obtaining an expression for the above, and it comes out to be:
\begin{equation}
	q(z) = -1 + \frac{3}{2} \frac{(1-\beta)(1-\nu)\tilde{\Omega}_m (1+z)^{3(1-\beta)(1-\nu)}}{\tilde{\Omega}_m (1+z)^{3(1-\beta)(1-\nu)} + \tilde{\Omega}_{\Lambda}}.
	\label{dec_para}
\end{equation}
The value of the deceleration parameter at the present time, $q_0$ is given by
\begin{equation}
	q_0 = -1 + \frac{3}{2}(1 - \beta)(1 - \Omega_\Lambda),
\end{equation}
In late times, as we approach the present epoch $z=0$, the deceleration parameter transitions from a positive to a negative value. This can be probed with another quantity called the transitional redshift. The expression for it is given by
\begin{equation}
	z_\text{tr} = -1 + \left[ \frac{2 (\Omega_\Lambda - \nu)}{(1 - \Omega_\Lambda)\left(3(1 - \beta)(1 - \nu) - 2\right)} \right]^{\frac{1}{3(1 - \beta)(1 - \nu)}}.
	\label{trans_redshift}
\end{equation}
Another key quantity that helps us understand the dynamics and the evolution of content in the Universe is the effective equation of state parameter $w_{\text{eff}}$. This work assumes that the decaying vacuum has an EoS parameter equal to $-1$. We will now compute the model's effective EoS, which includes both decaying dark energy and modified dark matter. This is calculated by the expression:
\begin{equation}
	w_{\text{eff}} = -1 - \frac{2}{3} \frac{\dot{H}}{H^2}.
	\label{eff_Eos2}
\end{equation}
On substituting the relevant values into \eqref{eff_Eos2}, we determine the effective EoS parameter to be as follows:
\begin{equation}
	w_{\text{eff}}(z) = -1 + \frac{(1-\beta)(1-\nu)\tilde{\Omega}_m (1+z)^{3(1-\beta)(1-\nu)}}{\tilde{\Omega}_m (1+z)^{3(1-\beta)(1-\nu)} + \tilde{\Omega}_{\Lambda}}.
	\label{eff_EoSpara}
\end{equation}
The present value of $w_\text{eff}$ which we denote by $w_0^{\text{eff}}$, is given by
\begin{equation}
	w_0^{\text{eff}} = -1 + (1 - \beta)(1 - \nu)\tilde{\Omega_m}.
\end{equation}

\section{Perturbations in the Cosmic Background}
\label{sec4}
This section gives a general overview of cosmic perturbation theory in the context of a matter creation cosmology with a decaying vacuum. Studying cosmic perturbations is crucial, as they underpin the formation of large-scale structures in the Universe—such as stars, galaxies, clusters, and quasars. The reader is referred to~\cite{Bran83,Muk92,Malik09,Sola18} for a detailed discussion on the theory of cosmological perturbations and the derivation of the perturbation equations. Here, we present only the basic equations and terms that are appropriate and relevant to our discussion.\\
\indent The differential equation for the matter density contrast, defined as $\delta_m \equiv \delta \rho_m / \rho_m$, in the considered model can be approximated as follows ~\cite{Perez24}
\begin{equation}
	\delta_m'' + \left( 3 + a \frac{H'(a)}{H(a)} \right) \frac{\delta_m'}{a} - \frac{3}{2} \Omega_m(a) \frac{\delta_m}{a^2} = 0,
	\label{pert_eq}
\end{equation}
where `prime' denotes differentiation with respect to the scale factor $a$. This second-order differential equation describes the evolution of the density contrast, as the primary effects arise from the specific form of the Hubble function.\\
\indent To solve \eqref{pert_eq}, we use the initial condition given in Ref.~\cite{daSilva21} as
\begin{equation}
	\delta_{mi} = 1.5 \times 10^{-4} \quad \text{at} \quad a_i = 10^{-3}, \quad \text{and} \quad \delta_{mi}' = \delta_{mi}/a_i.
	\label{pert_eq_ics}
\end{equation}
The expression for $\Omega_m(a)$ in terms of the model parameters can be derived through $\Omega_m(a) = 1 - \Omega_{\Lambda}(a)$, where
\begin{equation}
    \Omega_{\Lambda}(a) = \frac{\rho_{\Lambda}}{3 H^2 / 8 \pi G} = \frac{\Omega_{\Lambda}(a=1)-\nu}{{E(a)}^2} + \nu.
    \label{frac_density_para}
\end{equation}
Here $E(a)=H(a)/H_0$ is the dimensionless Hubble parameter in terms of the scale factor $a={(1+z)}^{-1}$. We substitute the required values into \eqref{pert_eq}, allowing us to numerically compute the matter density contrast $\delta_{mi}$ using the initial conditions \eqref{pert_eq_ics}. Next, we consider the linear growth rate of $\delta_{mi}$. This quantity is denoted by $f$, and in the linear theory, it is related to the peculiar velocity~\cite{Peeb93}. We define this as
\begin{equation}
	f(a) = \frac{d \ln D_m(a)}{d \ln a},
	\label{lineargrowthrate}
\end{equation}
where $D_m(a) = \frac{\delta_m(a)}{\delta_m(a = 1)}$ represents the linear growth function. The quantity denoted by $f\sigma_8$, called the `weighted linear growth rate', is obtained by multiplying the linear growth rate $f(z)$, defined in \eqref{lineargrowthrate}, and $\sigma_8(z)$. Here, $\sigma_8$ is a parameter that measures the density root-mean-square fluctuation on $8h^{-1}$ Mpc scales~\cite{Song09,Hute15}. It is defined~\cite{Ness08} as:
\begin{equation}
	\sigma_8(z) = \frac{\delta_m(z)}{\delta_m(z = 0)}\sigma_8(z = 0).
	\label{sigma8}
\end{equation}
Using \eqref{lineargrowthrate} and \eqref{sigma8}, the weighted linear growth rate is computed by the following relation:
\begin{equation}
	f\sigma_8(z) = -(1 + z) \frac{\sigma_8(z = 0)}{\delta_m(z = 0)} \frac{d\delta_m}{dz}.
	\label{fsigma8}
\end{equation}

\section{Data and Methodology}
\label{sec5}
This section considers the well-established and most recent observational samples to constrain the model parameters. We use the Bayesian Markov Chain Monte Carlo (MCMC) method for statistical analysis, which is conducted using the {\it emcee} package in Python. Optimization of the best fit of parameters is done by maximization of the probability function given by
\begin{equation}
\mathcal{L} \propto \exp(-\chi^2/2),
\end{equation}
where $\chi^2$ is the chi-squared function. We enlist the $\chi^2$ function for different data samples in the following subsections.
\subsection{Sample of Type Ia supernovae (SNIa)}
\label{sec5.1}
\noindent Often referred to as ``standard candles'', SNIa are among the most popular cosmological probes used to study the Universe's evolution. These supernovae are extremely luminous, with their peak brightness often rivalling that of their host galaxy~\cite{Leib00}. Increasingly precise data from SNIa have solidified the original ground-breaking observations and their implications for the expansion of the Universe~\cite{Riess04,Vasey07,Conley10,Suzuki12,Betoule14}.\\
\indent The comprehensive Pantheon sample, which includes luminosity distance measurements of 1048 supernovae across a redshift range of $0.01 < z < 2.26$, is used in this analysis. The observations allow us to constrain cosmological models using the observed distance modulus, $\mu_\text{obs}$. The theoretical distance modulus is defined as
\begin{equation}
	\mu_\text{th}(z, p) = 5 \log_{10} \left( \frac{D_L(z_\text{hel}, z_\text{cmb})}{1~\text{Mpc}} \right) + 25,
\end{equation}
where $p$ represents the parameter space, which varies across different cosmological models, and $D_L$ is the luminosity distance, expressed as
\begin{equation}
	D_L(z_\text{hel}, z_\text{cmb}) = (1 + z_\text{hel}) r(z_\text{cmb}).
\end{equation}
The comoving distance $r(z)$ is calculated as
\begin{equation}\label{comoving_distance}
	r(z) = c H_0^{-1} \int_{0}^{z} \frac{dz'}{E(z', p)},
\end{equation}
where $E(z) = H(z)/H_0$ is the dimensionless Hubble parameter, $c$ is the speed of light in vacuum, and $H_0$ is the Hubble constant. The CMB-frame and heliocentric redshifts are denoted by $z_\text{cmb}$ and $z_\text{hel}$, respectively.\\
\indent The Pantheon sample provides the observed distance modulus, $\mu_\text{obs}$, along with two separate components for uncertainty: a column for the statistical errors, which we reshape as a diagonal matrix $D_\text{stat}$, and a full systematic covariance matrix $C_\text{sys}$. These are combined to construct the total covariance matrix,
\begin{equation}
	C = D_\text{stat} + C_\text{sys}.
\end{equation}
which is used in the $\chi^2$ computation after taking the inverse.
Instead of fixing the absolute magnitude $M$, we analytically marginalize over it using the method described in~\cite{Conley10}, which avoids direct dependence on the nuisance parameter. Thus, we obtain the $\chi^2$ function for Pantheon SNIa as
\begin{equation}
	\chi^2_{\text{SNIa}} = a + \ln\left( \frac{e}{2\pi} \right) - \frac{b^2}{e},
\end{equation}
where $a = \Delta\mu^T C^{-1} \Delta\mu$, $b = \Delta\mu^T C^{-1} \mathbf{1}$, and $e = \mathbf{1}^T C^{-1} \mathbf{1}$, with $\Delta\mu = \mu_\text{th} - \mu_\text{obs}$ and $\mathbf{1}$ denoting a vector of ones.

\subsection{Baryon Acoustic Oscillations}
\label{sec5.2}
\noindent The measurement of the background expansion history of the Universe can be achieved through the standard ruler provided by the imprint of Baryon Acoustic Oscillations (BAO). The BAO method has been extensively studied and validated in numerous works ~\cite{Weinberg13,Eise05,Bass10}, providing a critical tool for constraining cosmological models. We present the basic equations and terms necessary for working with BAO measurements.\\
\indent The comoving sound horizon is given by
\begin{equation}\label{comoving_sound_horizon}
r_s(z) = \frac{c}{H_0} \int_{z}^{\infty} \frac{c_s}{E(z')} dz',
\end{equation}
where $c$ is the speed of light in vacuum and $c_s$ is the speed of sound, given by
\begin{equation}\label{speed_of_sound}
c_s = \frac{1}{\sqrt{3(1+R_b a)}},
\end{equation}
where $R_b$ gives the baryon to photon density ratio, calculated as
\begin{equation}
R_b a = \frac{3 \rho_b}{4 \rho_\gamma}, \quad R_b = 31500 \, \omega_b \left( \frac{T_{\text{CMB}}}{2.7 \text{ K}} \right)^{-4}.
\end{equation}
Here, $\omega_b = \Omega_b h^2$ is the baryon density parameter with $h = H_0/100$, and $T_{\text{CMB}}$ is the current measured value of the CMB temperature.\\
\indent One crucial period in our cosmic history is the baryon drag epoch, which refers to the time when baryons decoupled from photons. The redshift at drag epoch is calculated by the fitting function:
\begin{equation}\label{redshift_drag_epoch}
z_d = \frac{1345 \, \omega_m^{0.251}}{1 + 0.659 \, \omega_m^{0.828}} \left( 1 + b_1 \, \omega_b^{b_2} \right),
\end{equation}
where \( b_1 = 0.313 \, \omega_m^{-0.419} \left( 1 + 0.607 \, \omega_m^{0.674} \right) \) and \( b_2 = 0.238 \, \omega_m^{0.223} \).
Here $\omega_m = \Omega_m h^2$ is the matter density parameter.
By evaluating Eq. \eqref{comoving_sound_horizon} at the drag epoch redshift $z_d$, we obtain a key quantity $r_d$ which is a crucial term in BAO measurements. This gives the maximum distance sound waves could travel in the early Universe till the time of the last scattering (before the photon-baryon decoupling).\\
\indent We use the 2024 BAO measurements from the first year observations of the Dark Energy Spectroscopic Instrument (DESI). The complete data, along with the survey details, can be found in Ref.~\cite{DESI2024III,DESI2024VI}. We constrain the model parameters from the BAO data using three observables defined below, within the redshift range $0.295 \leq z \leq 2.330$.\\
\noindent The first observable is the comoving angular distance $D_M(z)$ which is given by
\begin{equation}
	D_M(z) = \int_{0}^{z} \frac{c}{H(z')} \, dz'.
	\label{angdist}
\end{equation}
Then we have the Hubble distance parameter $D_H(z)$, which is given by
\begin{equation}
	D_H(z) = \frac{c}{H(z)}.
	\label{hubbledist}
\end{equation}
From the former two, we calculate the angle-average distance, which is given by
\begin{equation}
	D_V(z) = (z {D_M(z)}^2 D_H(z))^{1/3}.
	\label{angavgdist}
\end{equation}
These quantities are normalized by the sound horizon at drag epoch $r_d$, which provides a standard ruler for measuring on large scales. Thus, we calculate our theoretical values against the observed values $D_M/r_d, D_H/r_d$ and $D_V/r_d$ obtained from the BAO measurements. The $\chi^2$ functions for the BAO data are:
\begin{equation}
	\chi^2_{D_M/r_d} = \sum_{i=1}^{5} \frac{\left( D_M(z_i, p) - D_{M, \text{obs}}(z_i) \right)^2}{\sigma_{D_M}^2(z_i)},
	\label{chi2BAO1}
\end{equation}
\begin{equation}
	\chi^2_{D_H/r_d} = \sum_{i=1}^{5} \frac{\left( D_H(z_i, p) - D_{H, \text{obs}}(z_i) \right)^2}{\sigma_{D_H}^2(z_i)},
	\label{chi2BAO2}
\end{equation}
\begin{equation}
	\chi^2_{D_V/r_d} = \sum_{i=1}^{2} \frac{\left( D_V(z_i, p) - D_{V, \text{obs}}(z_i) \right)^2}{\sigma_{D_V}^2(z_i)},
	\label{chi2BAO3}
\end{equation}
 where $D_M(z_i, p), D_H(z_i, p)$ and $D_V(z_i, p)$ are the theoretical values of the observable based on the model, and $D_{M, \text{obs}}$, $D_{H, \text{obs}}$ and $D_{V, \text{obs}}$ are the observed values of the DESI BAO measurements.\\
\indent Thus, the total $\chi^2$ value for the BAO data is
\begin{equation}
	\chi^2_{BAO} = \chi^2_{D_M/r_d} + \chi^2_{D_H/r_d} + \chi^2_{D_V/r_d}.
	\label{chi2BAO}
\end{equation}

\subsection{Cosmic Microwave Background}
\label{sec5.3}
\noindent The Cosmic Microwave Background (CMB) measurements of the locations and amplitudes of peaks of acoustic oscillations give information about the distance to the surface of last scattering, given by $r(z_*)$. This can be evaluated through Eq.~\eqref{comoving_distance} at $z_*$, which is the redshift at the time of recombination or last scattering. It marks the epoch when photons last scattered off free electrons before being able to freely travel through the Universe, leaving behind their imprint in the form of the CMB radiation. The reader is referred to~\cite{Gawi00,Fixsen09} for a more detailed discussion on CMB. We now proceed with the basic equations employed in the data analysis.\\

\noindent The fitting function for recombination redshift $z_*$ is
\begin{equation}\label{redshift_recombination}
z_* = 1048 \left( 1 + 0.00124 \, \omega_b^{-0.738} \right) \left( 1 + g_1 \, \omega_m^{g_2} \right),
\end{equation}
where
\begin{equation}
g_1 = \frac{0.0783 \, {\omega_b}^{-0.238}}{1 + 39.5 \, \omega_b^{0.763}}, \quad
g_2 = \frac{0.560}{1 + 21.1 \, \omega_b^{1.81}}.
\end{equation}
\indent Here $\omega_m$ and $\omega_b$ are the matter and baryon density parameters, respectively. We are particularly interested in two quantities that characterize the CMB spectrum: the shift parameter and acoustic length scale.\\
\indent It is observed that through the acoustic oscillations, Doppler peaks are produced in the photon (radiation) spectrum. The placement of the peaks in the CMB spectrum is affected by the matter and radiation density, along with the dark energy. The shift in these peaks is probed by the shift parameter $\mathcal{R}$. The spectrum is affected in the line-of-sight direction, leading to the peak heights. We obtain $\mathcal{R}$ through the relation:
\begin{equation}\label{shift_parameter}
  \mathcal{R} = \sqrt{\Omega_m} H_0 \frac{r(z_*)}{c}.
\end{equation}
The acoustic scale $l_a$ determines the spacing between the peaks and characterizes the CMB temperature power spectrum in the transverse direction. This is calculated by
\begin{equation}\label{acoustic_scale}
l_A = \frac{\pi r(z_*)}{r_s(z_*)}.
\end{equation}
\noindent When working with CMB, we have to consider the radiation component, as in early times it is dominant and its effect cannot be neglected. We use the relation $\Omega_r = \Omega_m / (1 + z_{\text{eq}})$ to determine the radiation density, where $z_{\text{eq}}$ is the epoch of matter-radiation equality and is given by
\begin{equation}\label{z_equality}
  z_{\text{eq}} = 2.5 \times 10^4 \, \omega_m \left( \frac{T_{\text{CMB}}}{2.7\,\text{K}} \right)^{-4},
\end{equation}
where $T_{\text{CMB}}$ represents the present-day temperature of the CMB. We adopt the best-fit value $T_{\text{CMB}} = 2.7255 \, \text{K} $, measured by the COBE-FIRAS instrument~\cite{Fixsen96}.\\
\indent CMB observation contribution in the constraint of cosmological parameters can be converted to CMB shift parameters for breaking degeneracy of parameters, instead of using the entire CMB power spectrum.~\cite{Wang07,Wang13}  This allows us to consider the data vector $\mathbf{v} = (R, l_A, \omega_b)^T$ which summarizes the Planck 2018 TT,TE,EE+lowE results~\cite{Tris24}. The vector $\mathbf{v}$, along with the covariance matrix $C_{\mathbf{v}}$ have been reported in~\cite{Liu24}. The distance priors used in our analysis are obtained assuming a $\Lambda$CDM framework. However, since the additional parameters in our model, when set to zero, recover the $\Lambda$CDM limit, the use of these priors remains valid as an approximation. This ensures consistency at the background level during the early Universe.\\
\indent The $\chi^2$ function for the CMB data is given by:
\begin{equation}\label{chi2_CMB}
\chi^2_{\text{CMB}} = (\mathbf{v} - \mathbf{v}_{\text{th}})^T C_\mathbf{v}^{-1} (\mathbf{v} - \mathbf{v}_{\text{th}}),
\end{equation}
where $\mathbf{v}_{\text{th}}$ is the vector for the theoretical values of $R, l_A, \omega_b$; and $C_\mathbf{v}^{-1}$ is the inverse of the covariance matrix.

\subsection{Sample of Cosmic Chronometer (CC)}
\label{sec5.4}
\noindent The cosmic chronometer (CC) approach provides measurements of the Hubble parameter based on the galaxy differential age method using the largest galaxies that are passively evolving~\cite{More22}. In our analysis, we use 32 CC data points from different surveys that span a redshift range of $0.07 \leq z \leq 1.965$. The $\chi^2$ function for CC data is given by
\begin{equation}
	\chi^2_{CC} = \sum_{i=1}^{32} \frac{\left[ H(z_i, p) - H_\text{obs}(z_i) \right]^2}{\sigma_H^2(z_i)},
\end{equation}
where $H(z_i, p)$ represents the theoretical value of the Hubble parameter for the model parameters, $H_\text{obs}(z_i)$ is the observed value of the Hubble parameter at redshift $z_i$, and $\sigma_H(z_i)$ is the associated uncertainty with $H_\text{obs}(z_i)$.

\subsection{Weighted Linear Growth Rate Data}
\label{sec5.5}
We introduced the background evolution of matter perturbations in Section~\ref{sec4} and used Eq.~\eqref{fsigma8} to define the weighted linear growth rate. To provide a more comprehensive analysis of the present model in the context of perturbation evolution, we focus on the observable quantity $f(z)\sigma_8(z)$. For this purpose, we use a compilation of 26 measurements of $f(z)\sigma_8(z)$ obtained from various galaxy surveys given in Table 2 of Ref.~\cite{Bessa22}. We take $\sigma_8$ as a free parameter and constrain its value using the sample.\\
\indent We compare the observational data with the theoretical values calculated from our models using the $\chi^2$ function for $f(z)\sigma_8(z)$, which is defined as
\begin{equation}
	\chi^2_{f\sigma_8} = \sum_{i=1}^{18} \frac{\left[ f\sigma_{8, \text{theo}}(z_i, p) - f\sigma_{8, \text{obs}}(z_i) \right]^2}{\sigma^2_{f\sigma_8}(z_i)},
\end{equation}
where $f\sigma_{8, \text{theo}}(z_i, p)$ is the theoretical value computed using \eqref{fsigma8}, and $f\sigma_{8, \text{obs}}(z_i)$ represents the observed data.

\subsection{$H_0$ Prior}
\label{sec5.6}
Recent observations have revealed a notable tension (of around 4.89$\sigma$) between the value of the Hubble constant $H_0$ inferred from early-Universe probes (such as Planck CMB data) and that measured directly from late-Universe observations, most notably by the SH0ES collaboration. To explore whether our model can alleviate this tension, we incorporate a Gaussian prior on $H_0$ centered around the R22 measurement~\cite{Riess22}:
\begin{equation}\label{h0prior}
	H_0 = 73.04 \pm 1.04~\text{km/s/Mpc},
\end{equation}
Including this prior allows us to test whether the model accommodates higher $H_0$ values preferred by local measurements, and potentially mitigates the tension in a statistically consistent way; an approach that has widely been adopted~\cite{Perez24}.\\

\subsection{Parameter Estimation}
\label{sec5.7}
\indent Using three different combinations of the data discussed above, we employ Bayesian MCMC methods to explore the parameter space of the proposed model described in Section~\ref{sec3}, using the \texttt{emcee} Python package~\cite{Fore13,Trot08}. We consider the following combinations of observational datasets to constrain the model parameters. The total $\chi^2$ function in each case is constructed by summing the individual contributions from different probes:
\begin{itemize}
	\item \textbf{BASE}: This dataset includes observations from Type Ia Supernovae (SNIa), Baryon Acoustic Oscillations (BAO), Cosmic Chronometers (CC), and the weighted linear growth rate data. It consists of 1118 independent data points. The total chi-squared function is given by:
\begin{equation}
    \chi^2_{\text{tot}} = \chi^2_{\text{SNIa}} + \chi^2_{\text{BAO}} + \chi^2_{H(z)} + \chi^2_{f\sigma_8}.
\end{equation}	
\item \textbf{BASE + $H_0$ Prior}: In this combination, we supplement the BASE dataset with an external Gaussian prior on the Hubble constant, $H_0$ as defined in \eqref{h0prior}. The total chi-squared function remains the same as in the above case, as the effect of the prior is captured by the chi-squared components of the independent datasets.
\item \textbf{BASE + CMB + $H_0$ Prior}: Here, we include the CMB component using the set of distance priors and retain the $H_0$ prior as well. The effective data points are now 1121, and the total $\chi^2$ function becomes:
\begin{equation}
	\chi^2_{\text{tot}} = \chi^2_{\text{SNIa}} + \chi^2_{\text{BAO}} + \chi^2_{\text{CMB}} + \chi^2_{H(z)} + \chi^2_{f\sigma_8}.
\end{equation}
\end{itemize}
\indent Our parameter space consists of the parameters: $H_0, \Omega_\Lambda, \nu, \beta, \sigma_8$ across the three datasets with the additional parameter $\omega_b$ for the CMB case. We adopt flat (uniform) priors for the primary cosmological parameters, and weak Gaussian priors for the model-specific parameters $\nu$ and $\beta$, informed by preliminary runs. For the MCMC sampling, we employ at least 100 walkers and run the sampler for 2500 steps, discarding the initial 20\% as burn-in. Convergence of the chains is assessed using the Gelman-Rubin diagnostic~\cite{Gel92}. The figures are generated using the \texttt{getdist} and \texttt{matplotlib} libraries in Python.

\begin{table}[ht]
\centering
\begin{tabular}{|c|cc|cc|}
\hline
\textbf{Parameter} & \multicolumn{2}{c|}{\textbf{BASE}} & \multicolumn{2}{c|}{\textbf{BASE + $H_0$}} \\
\cline{2-5}
 & $\boldsymbol{\Lambda}$\textbf{CDM} & $\boldsymbol{\Gamma}$\textbf{RVM} & $\boldsymbol{\Lambda}$\textbf{CDM} & $\boldsymbol{\Gamma}$\textbf{RVM} \\
\hline
$H_0$ [km/s/Mpc] & $68.84^{+0.88}_{-0.89}$ & $68.67^{+0.87}_{-0.86}$ & $70.64^{+0.67}_{-0.68}$ & $70.53^{+0.67}_{-0.68}$ \\
$\Omega_\Lambda$ & $0.698^{+0.015}_{-0.016}$ & $0.689^{+0.016}_{-0.017}$ & $0.724 \pm 0.011$ & $0.717^{+0.012}_{-0.013}$ \\
$\nu$ & -- & $0.00218^{+0.00141}_{-0.00124}$ & -- & $0.00209^{+0.00138}_{-0.00122}$ \\
$\beta$ & -- & $0.00628^{+0.00378}_{-0.00349}$ & -- & $0.00579^{+0.00363}_{-0.00333}$ \\
$\sigma_8$ & $0.802^{+0.022}_{-0.023}$ & $0.802^{+0.024}_{-0.023}$ & $0.818 \pm 0.023$ & $0.818 \pm 0.023$ \\
$S_8$ & $0.805^{+0.025}_{-0.024}$ & $0.816 \pm 0.026$ & $0.784 \pm 0.023$ & $0.794^{+0.023}_{-0.024}$ \\
$q_0$ & $-0.547^{+0.024}_{-0.023}$ & $-0.537^{+0.024}_{-0.023}$ & $-0.587 \pm 0.017$ & $-0.579^{+0.019}_{-0.018}$ \\
$z_{\text{tr}}$ & $0.666 \pm 0.041$ & $0.663 \pm 0.041$ & $0.739^{+0.034}_{-0.032}$ & $0.739^{+0.034}_{-0.035}$ \\
$w_0^{\text{eff}}$ & $-0.698^{+0.016}_{-0.015}$ & $-0.691^{+0.016}_{-0.015}$ & $-0.724 \pm 0.011$ & $-0.719 \pm 0.012$ \\
$t_0$ [Gyr] & $13.67 \pm 0.09$ & $13.70 \pm 0.09$ & $13.66 \pm 0.09$ & $13.69 \pm 0.09$ \\
\hline
\end{tabular}
\caption{Parameter constraints with 68\% confidence limits (1-$\sigma$) for the cosmological parameters of the $\Lambda$CDM and $\Gamma$RVM models from the BASE and BASE+$H_0$ datasets. The BASE dataset contains the data: SNIa, H(z), BAO, and $f\sigma_8$, while the BASE+$H_0$ dataset includes an additional $H_0$ prior.}
\label{table1}
\end{table}

\begin{table}[ht]
\centering
\begin{tabular}{|c|cc|}
\hline
\textbf{Parameter} & \multicolumn{2}{c|}{\textbf{BASE + CMB + $H_0$}} \\
\cline{2-3}
 & $\boldsymbol{\Lambda}$\textbf{CDM} & $\boldsymbol{\Gamma}$\textbf{RVM} \\
\hline
$H_0$ [km/s/Mpc] & $68.68 \pm 0.39$ & $68.33 \pm 0.40$ \\
$\Omega_\Lambda$ & $0.702 \pm 0.005$ & $0.708 \pm 0.005$ \\
$\nu$ & -- & $0.00067^{+0.00036}_{-0.00034}$ \\
$\beta$ & -- & $0.00087^{+0.00045}_{-0.00042}$ \\
$\omega_b$ & $0.02262 \pm 0.00013$ & $0.02254 \pm 0.00013$ \\
$\sigma_8$ & $0.803 \pm 0.021$ & $0.809 \pm 0.022$ \\
$S_8$ & $0.801 \pm 0.021$ & $0.797^{+0.022}_{-0.021}$ \\
$r_d$ [Mpc] & $147.35^{+0.23}_{-0.22}$ & $149.96^{+0.89}_{-0.84}$ \\
$q_0$ & $-0.553 \pm 0.007$ & $-0.563 \pm 0.008$ \\
$z_{\text{tr}}$ & $0.677 \pm 0.013$ & $0.697 \pm 0.015$ \\
$w_0^{\text{eff}}$ & $-0.702 \pm 0.005$ & $-0.708 \pm 0.005$ \\
$t_0$ [Gyr] & $13.75 \pm 0.02$ & $13.92 \pm 0.06$ \\
\hline
\end{tabular}
\caption{Parameter constraints for the $\Lambda$CDM and $\Gamma$RVM model obtained from the BASE+CMB+$H_0$ data. The baryon density parameter $\omega_b$ and the sound horizon at the drag epoch $r_d$ have also been reported to probe early-universe physics.}
\label{table2}
\end{table}

\begin{figure*}[ht]
    \centering
    \includegraphics[width=0.75\textwidth]{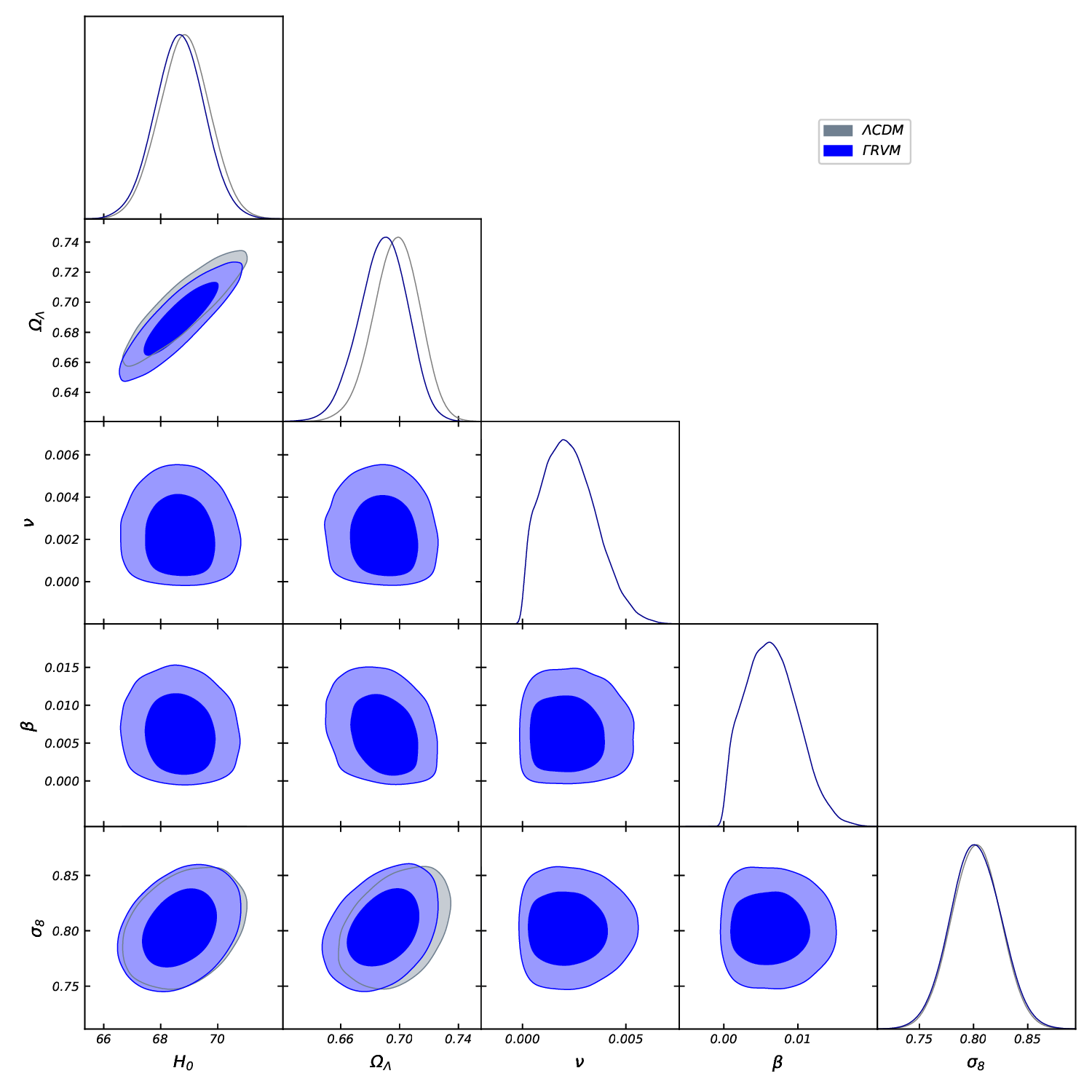}
    \caption{One-dimensional marginalized distributions and two-dimensional confidence contours (68.3\% and 95.4\%) for the $\Lambda$CDM and $\Gamma$RVM models using the BASE dataset: SNIa, H(z), BAO, and $f\sigma_8$).}
    \label{fig1}
\end{figure*}

\begin{figure*}[ht]
    \centering
    \includegraphics[width=0.75\textwidth]{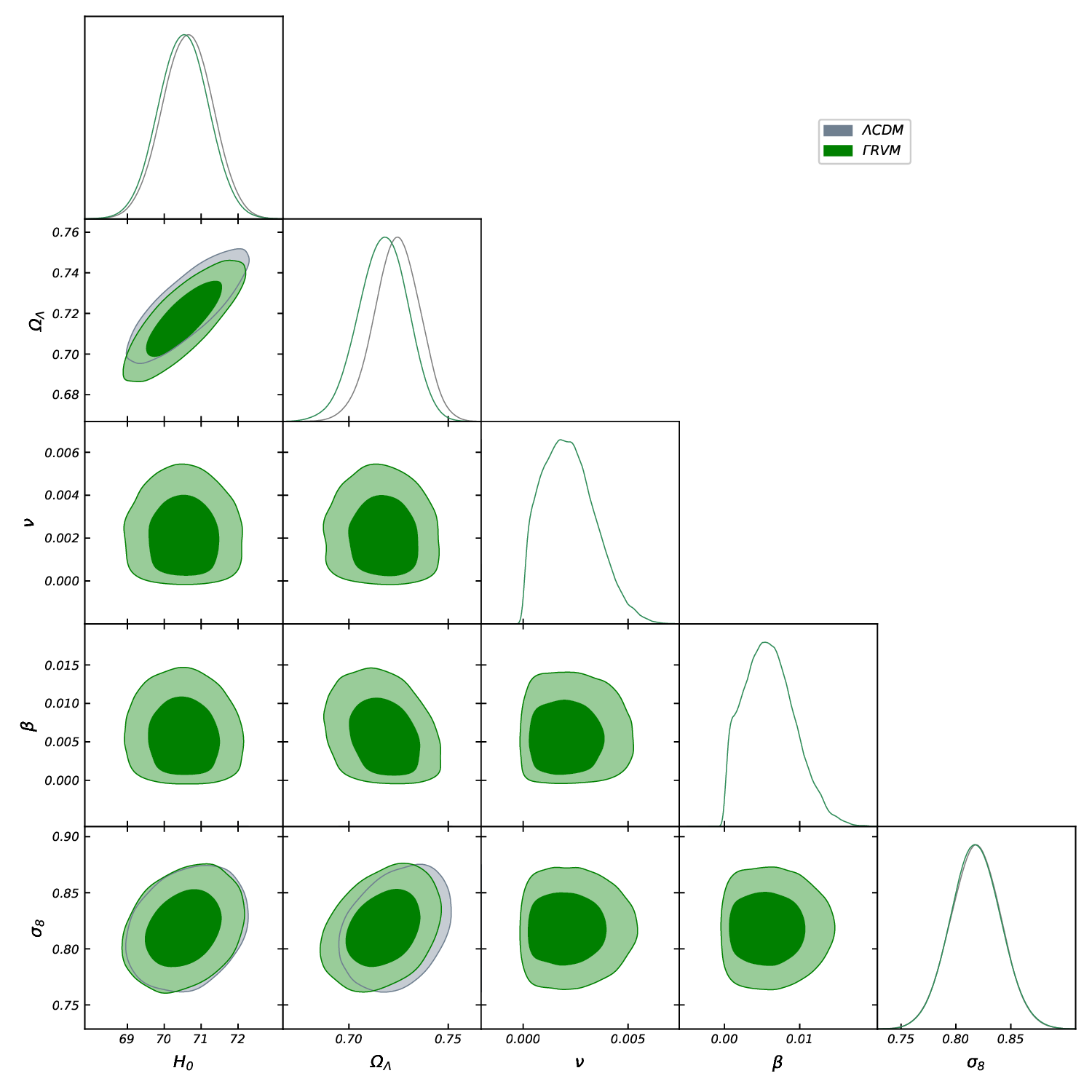}
    \caption{One-dimensional marginalized distributions and two-dimensional confidence contours obtained using the \textbf{BASE+$H_0$} data.}
    \label{fig2}
\end{figure*}

\begin{figure*}[ht]
    \centering
    \includegraphics[width=0.85\textwidth]{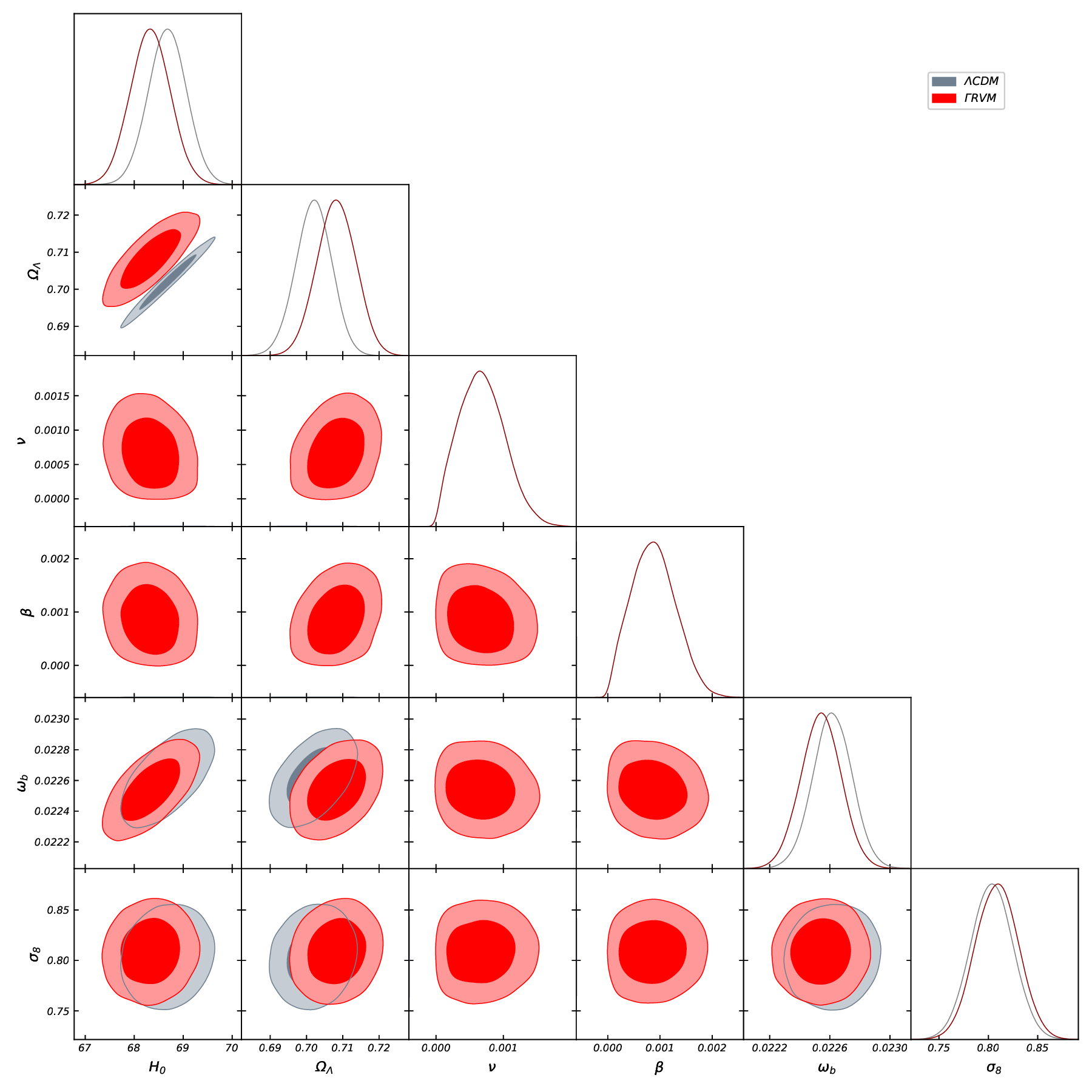}
    \caption{One-dimensional marginalized distributions and two-dimensional confidence contours with the inclusion of the CMB distance priors.}
    \label{fig3}
\end{figure*}

\begin{figure*}[ht]
    \centering
    \includegraphics[width=0.75\textwidth]{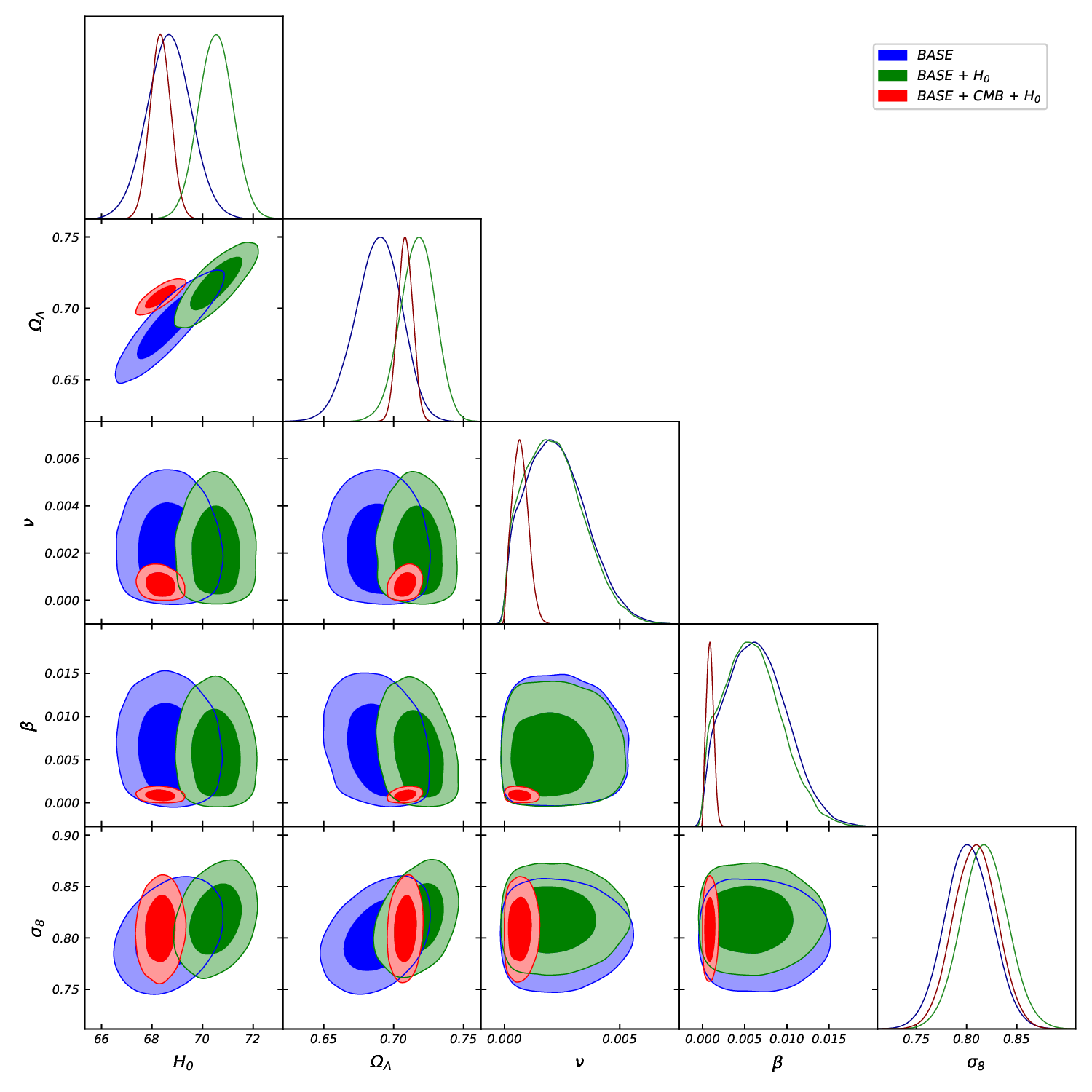}
    \caption{The combined likelihoods and confidence contours for the $\Gamma$RVM model across the three datasets. Clearly addition of CMB distance priors provides tighter constraints.}
    \label{fig4}
\end{figure*}

\begin{figure}[htbp]
\centering
	\begin{minipage}[b]{0.85\textwidth}
		%\centering
		\scalebox{0.4}{\includegraphics{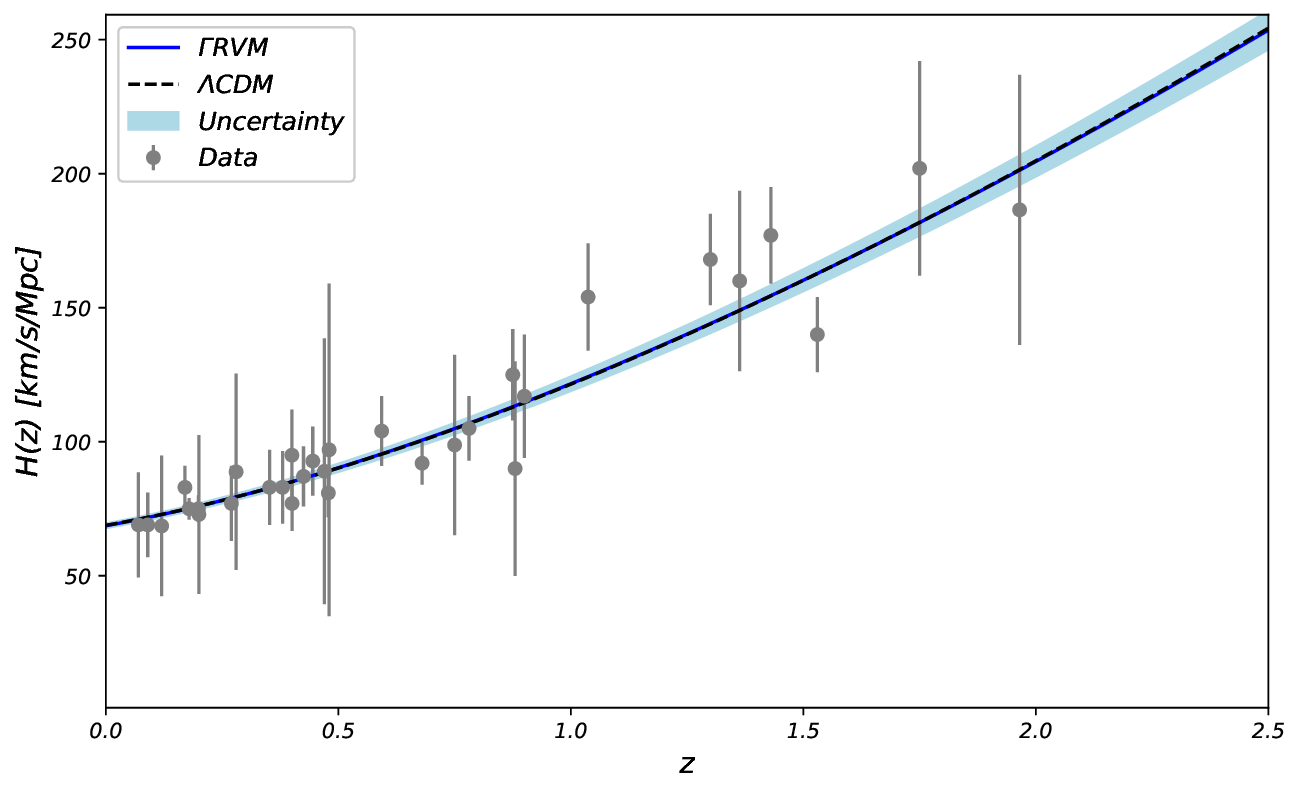}}
		\caption{Comparison of the 32 CC data points with the Hubble Parameter function $H(z)$ against redshift $z$ for the constrained parameter estimates of the $\Lambda$CDM (black dashed) and $\Gamma$RVM model (blue) obtained from BASE data. The shaded region represents the uncertainty for the latter and the grey points represent the data.}
		\label{fig5}
	\end{minipage}
    \begin{minipage}[b]{0.85\textwidth}
		%\centering
		\scalebox{0.4}{\includegraphics{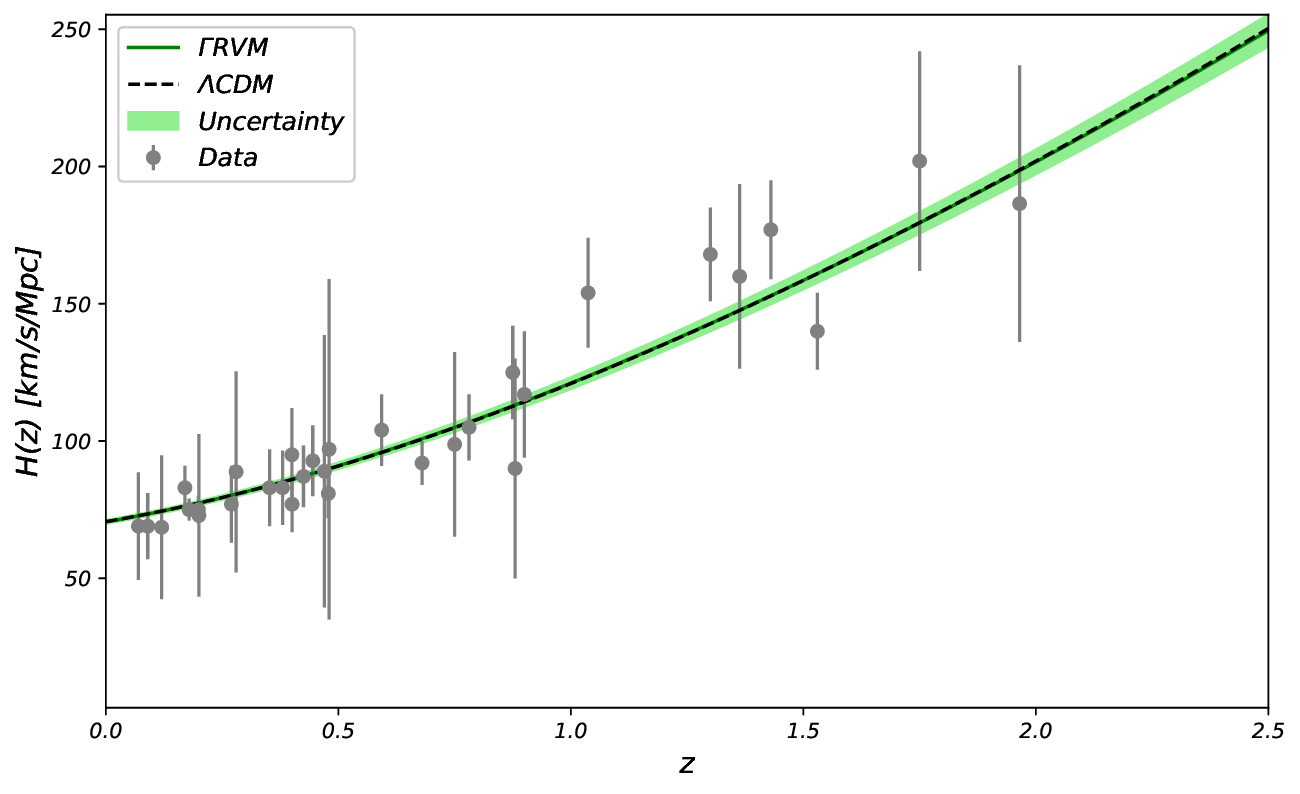}}
		\caption{Same as in Fig.~\ref{fig5}, but now we consider the estimates obtained from the BASE+$H_0$ data. The $H(z)$ trajectory for the model is represented by the green line.}
		\label{fig6}
	\end{minipage}
\begin{minipage}[b]{0.85\textwidth}
		%\centering
		\scalebox{0.4}{\includegraphics{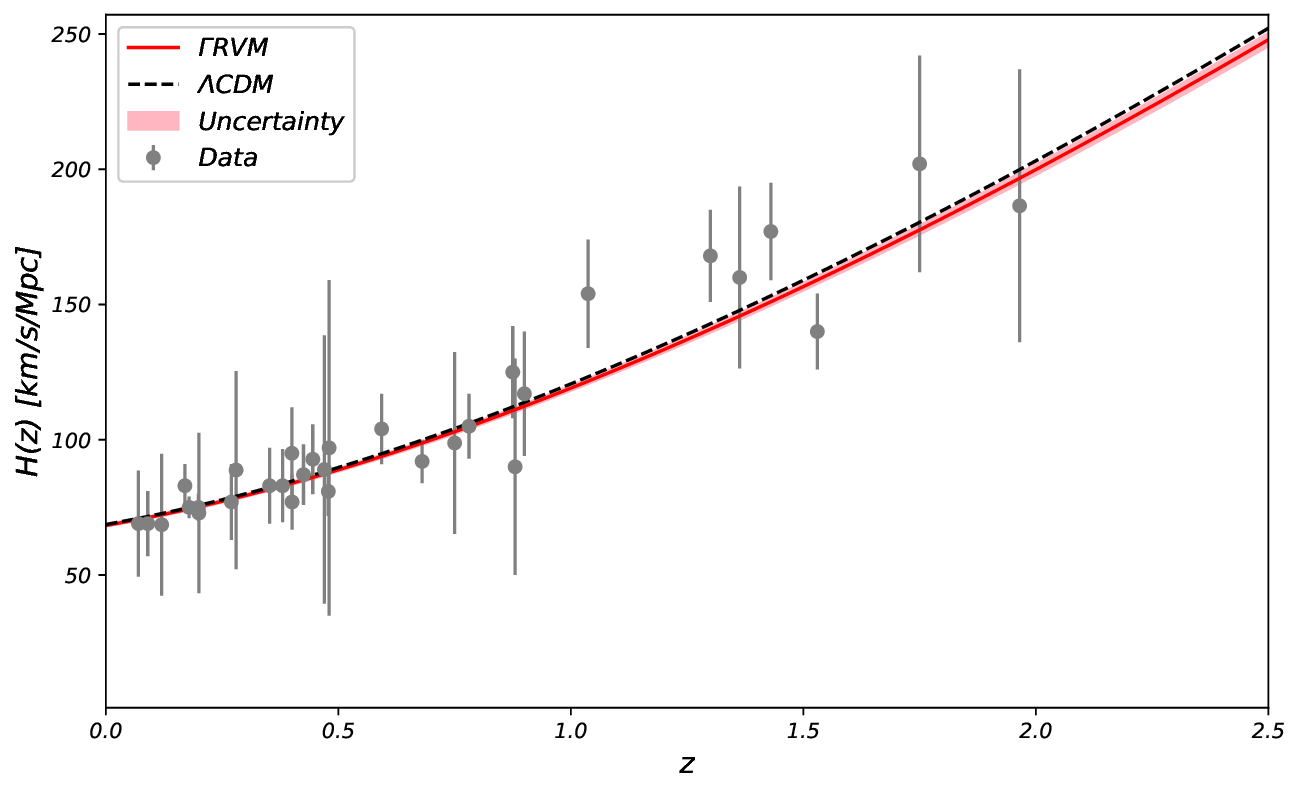}}
		\caption{Same as in Fig.~\ref{fig5} but now we consider the BASE+CMB+$H_0$ data. The $H(z)$ trajectory for the model is represented by the red line and we see that with inclusion of CMB, the uncertainty is narrower.}
		\label{fig7}
	\end{minipage}
\end{figure}

\begin{figure}[htbp]
\centering
	\begin{minipage}[b]{0.85\textwidth}
		%\centering
		\scalebox{0.4}{\includegraphics{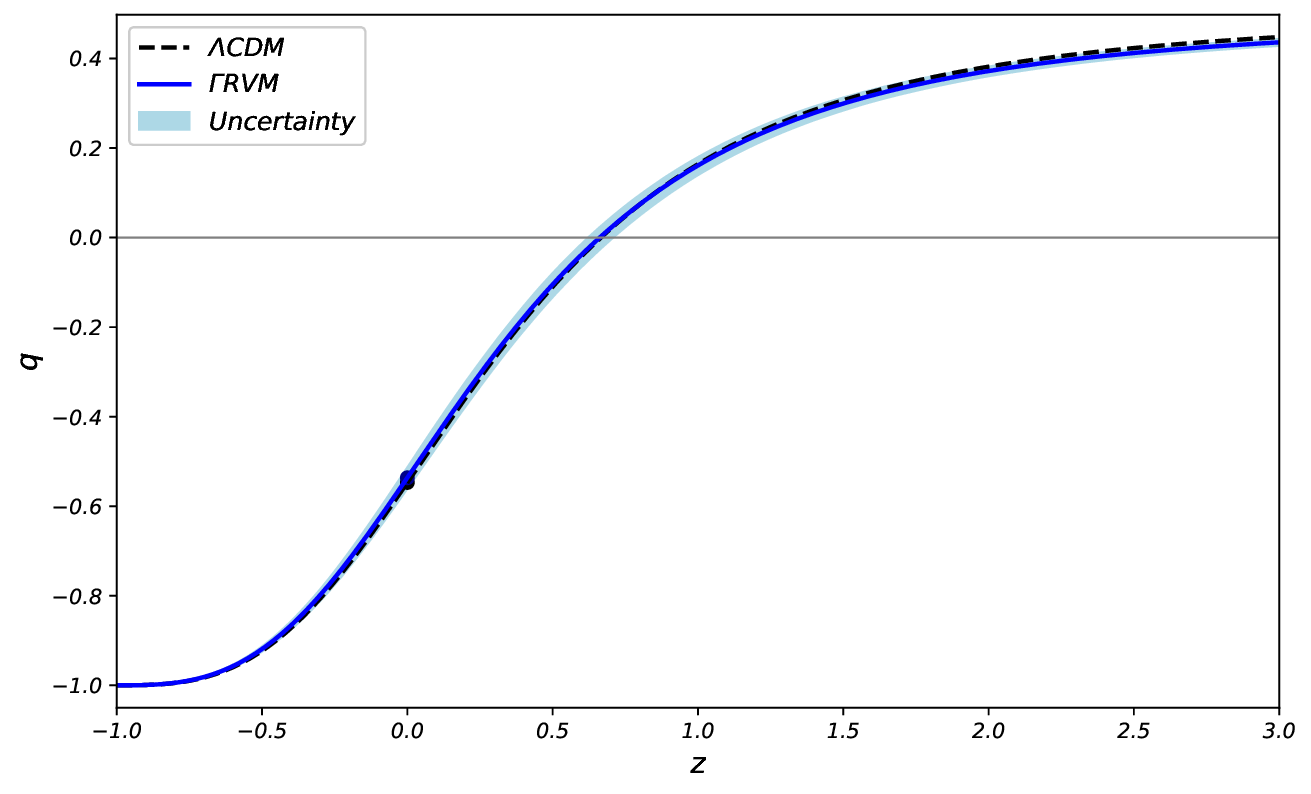}}
		\caption{Plots of deceleration parameter $q(z)$ against redshift $z$ for the $\Lambda$CDM and the $\Gamma$RVM model represented by the dashed black and solid blue line respectively. The 1-$\sigma$ uncertainty has been shown for the latter along with the present values for both marked by the solid dots. The parameter values are taken from the estimates reported in Table~\ref{table1}.}
		\label{fig8}
	\end{minipage}
	\begin{minipage}[b]{0.85\textwidth}
		%\centering
		\scalebox{0.4}{\includegraphics{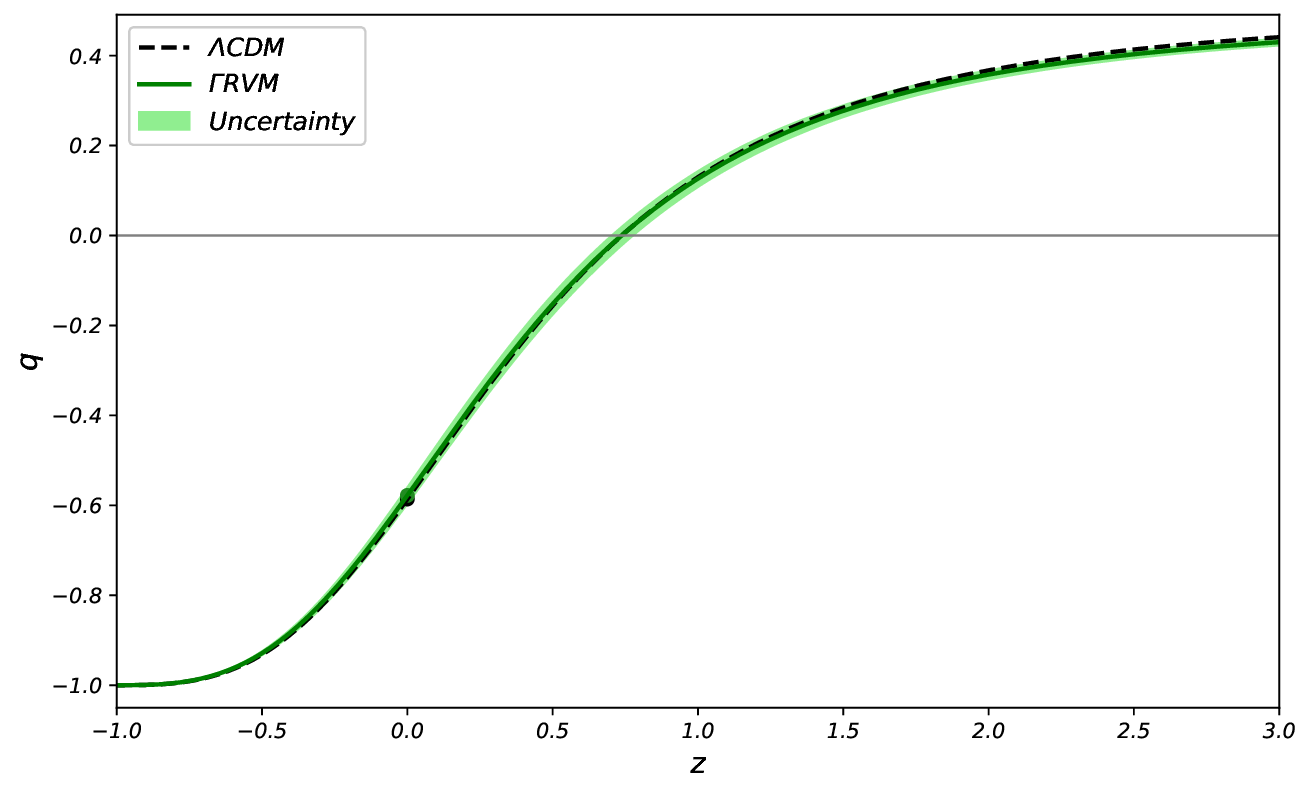}}
		\caption{Same as in Fig.~\ref{fig8} but with the $BASE+H_0$ data. The model trajectory is represented by the green line.}
		\label{fig9}
	\end{minipage}
	\begin{minipage}[b]{0.85\textwidth}
		%\centering
		\scalebox{0.4}{\includegraphics{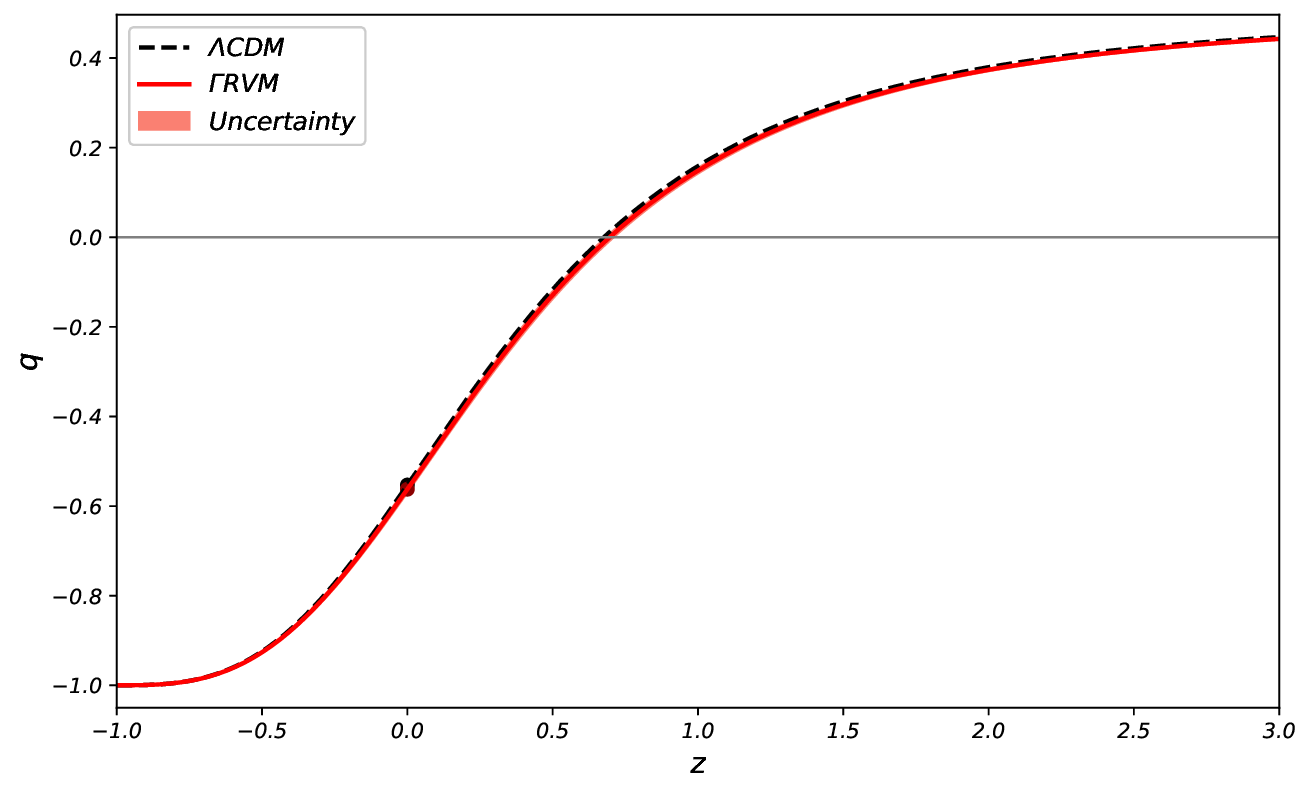}}
		\caption{Same as in Fig.~\ref{fig8} but with the $BASE+CMB+H_0$ data whose parameter estimates have been reported in Table~\ref{table2}. The model trajectory is represented by the red line.}
		\label{fig10}
	\end{minipage}
\end{figure}

\begin{figure}[htbp]
\centering
	\begin{minipage}[b]{0.85\textwidth}
		%\centering
		\scalebox{0.4}{\includegraphics{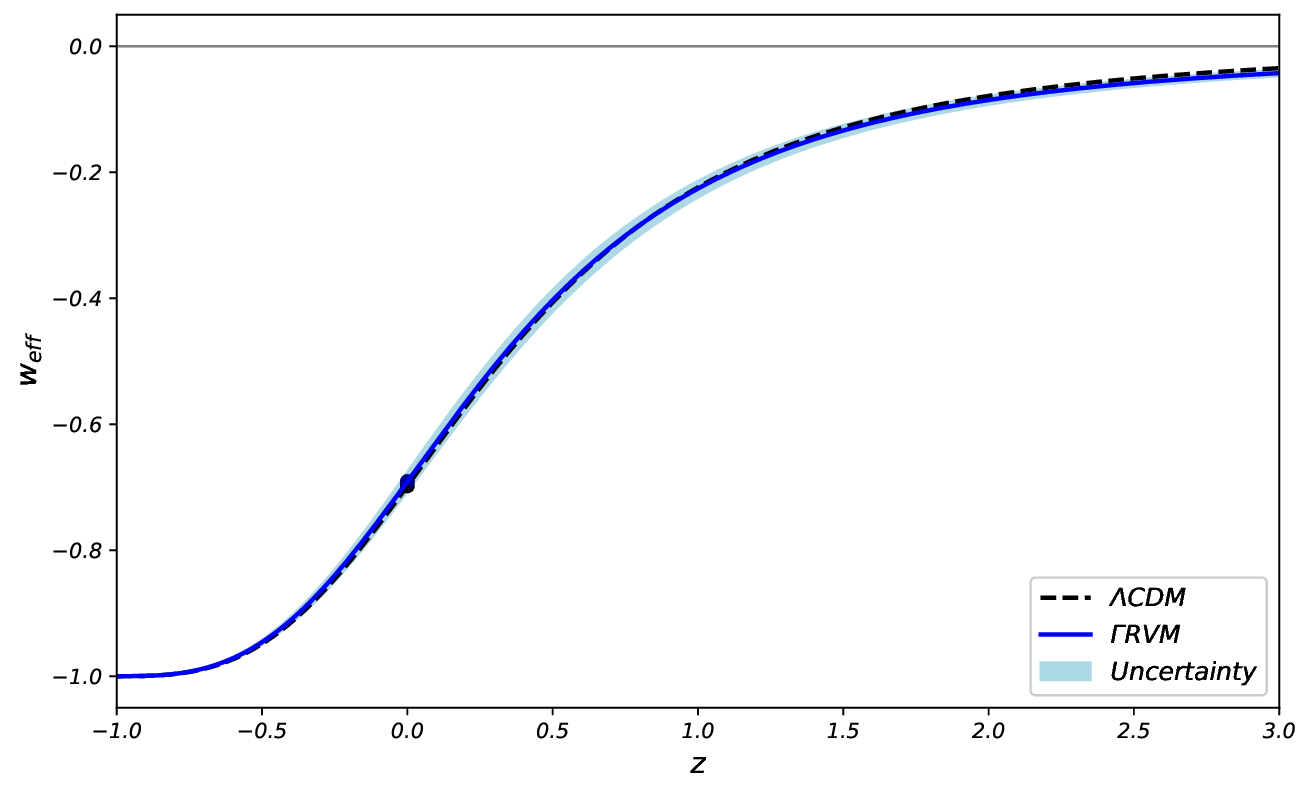}}
		\caption{Plots of effective EoS parameter $w_\text{eff}(z)$ with redshift $z$ for the standard model and the proposed model represented by the dashed black and solid blue line respectively. The 1-$\sigma$ uncertainty has been shown for the latter along with the present values for both marked by the solid dots. The parameter values are taken from the estimates reported in Table~\ref{table1}.}
		\label{fig11}
	\end{minipage}
	\begin{minipage}[b]{0.85\textwidth}
		%\centering
		\scalebox{0.4}{\includegraphics{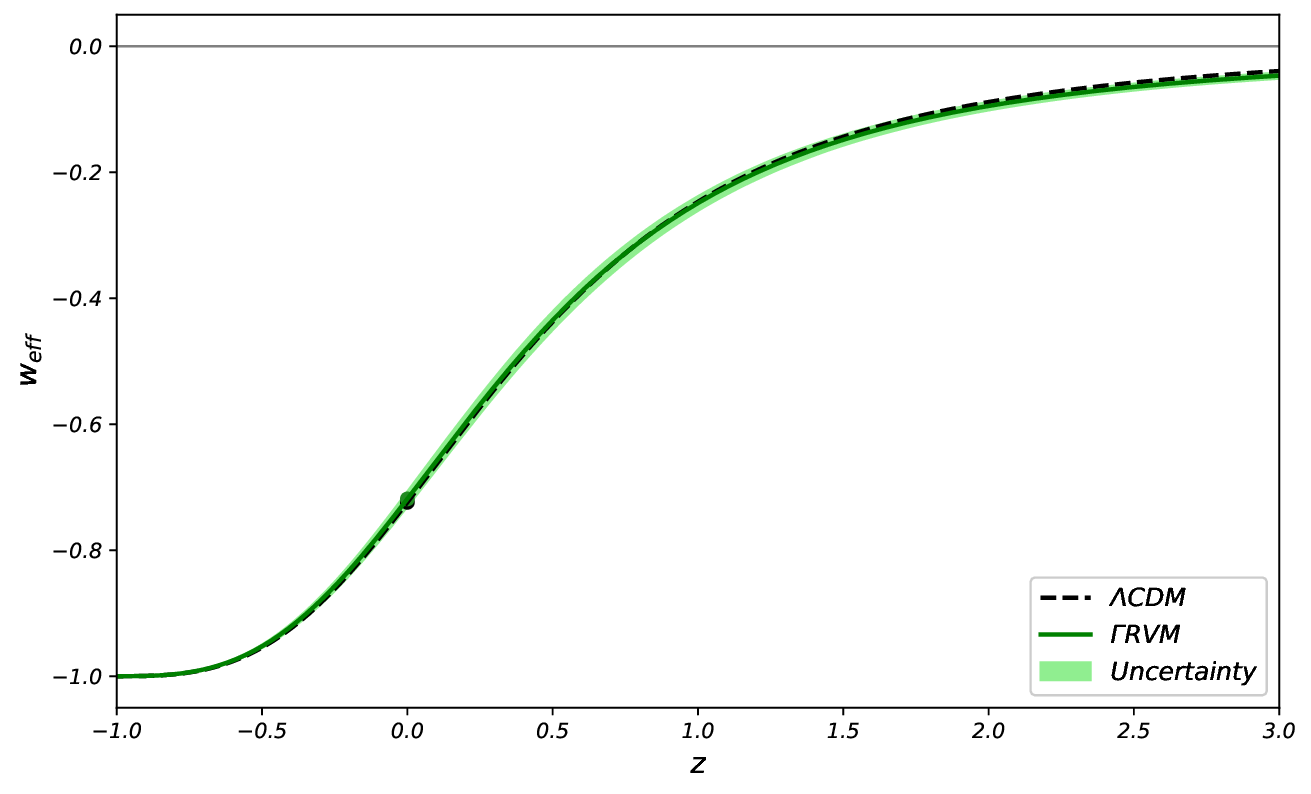}}
		\caption{Same as in Fig.~\ref{fig11} but with the $BASE+H_0$ data. The model trajectory is represented by the green line.}
		\label{fig12}
	\end{minipage}
	\begin{minipage}[b]{0.85\textwidth}
		%\centering
		\scalebox{0.4}{\includegraphics{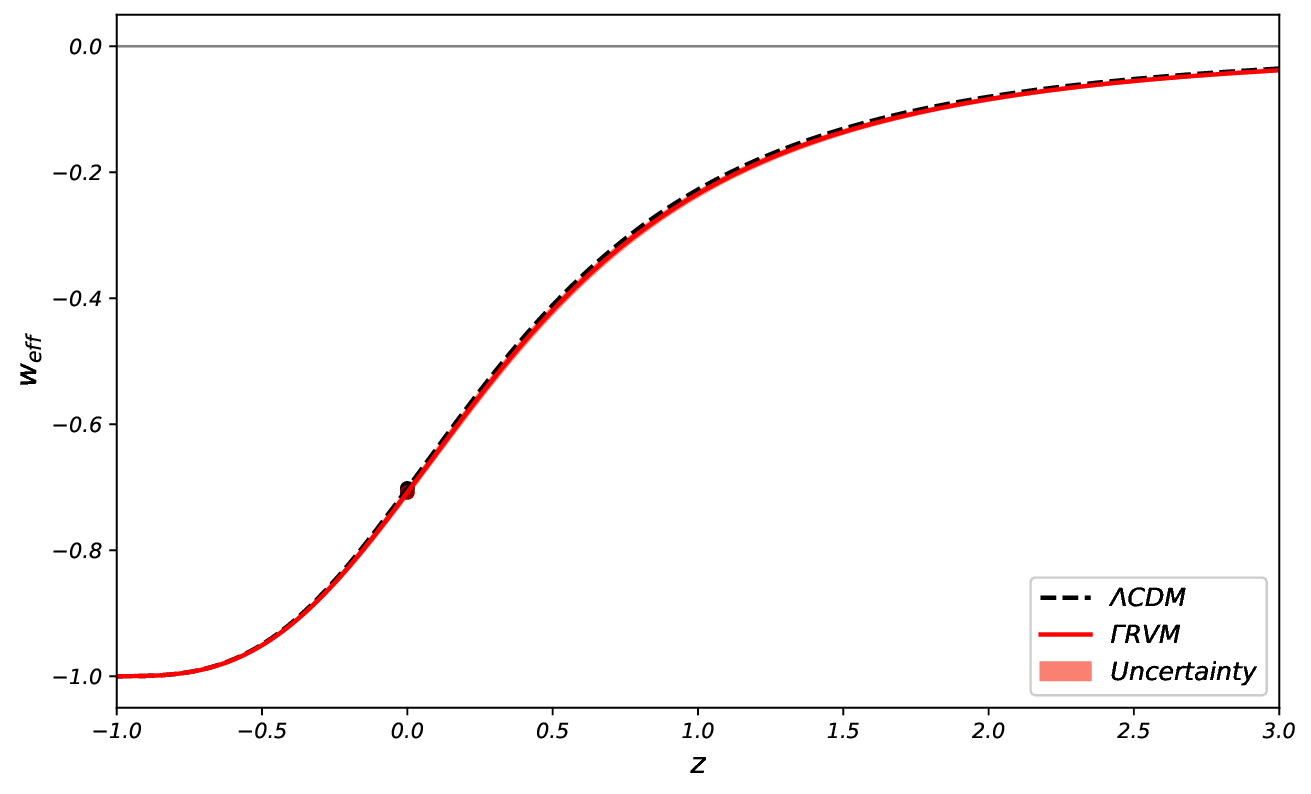}}
		\caption{Same as in Fig.~\ref{fig11} but with the $BASE+CMB+H_0$ data whose parameter estimates have been reported in Table~\ref{table2}. The model trajectory is represented by the red line.}
		\label{fig13}
	\end{minipage}
\end{figure}

\begin{figure}[htbp]
\centering
	\begin{minipage}[b]{0.85\textwidth}
		%\centering
		\scalebox{0.38}{\includegraphics{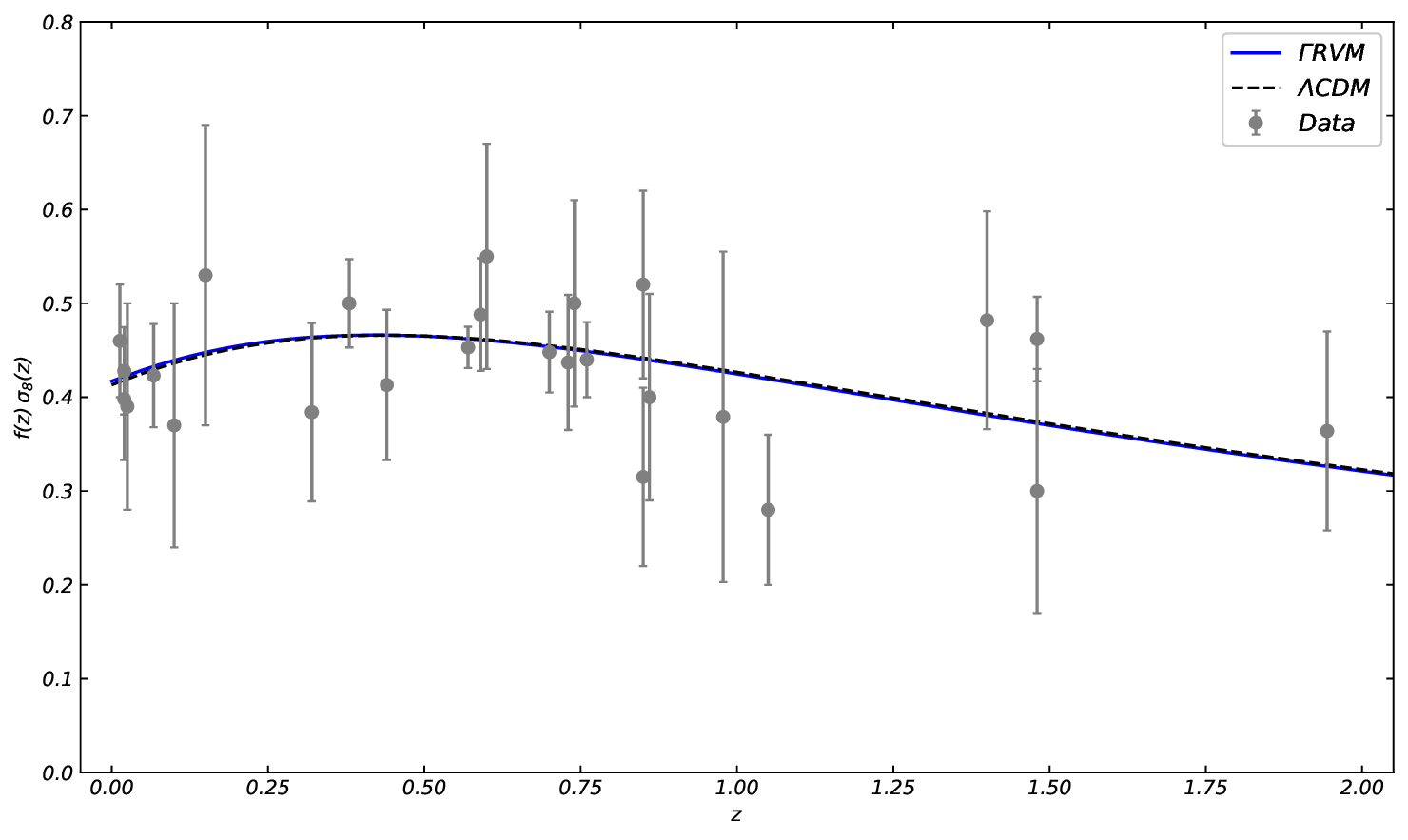}}
		\caption{Trajectories for the weighted linear growth rate $f(z)\sigma_8(z)$, shown for the standard model (black dashed) and the $\Gamma$RVM model (blue). The curves are derived using the parameter constraints obtained from the BASE dataset, given in Table~\ref{table1}.}
		\label{fig14}
	\end{minipage}
	\begin{minipage}[b]{0.85\textwidth}
		%\centering
		\scalebox{0.38}{\includegraphics{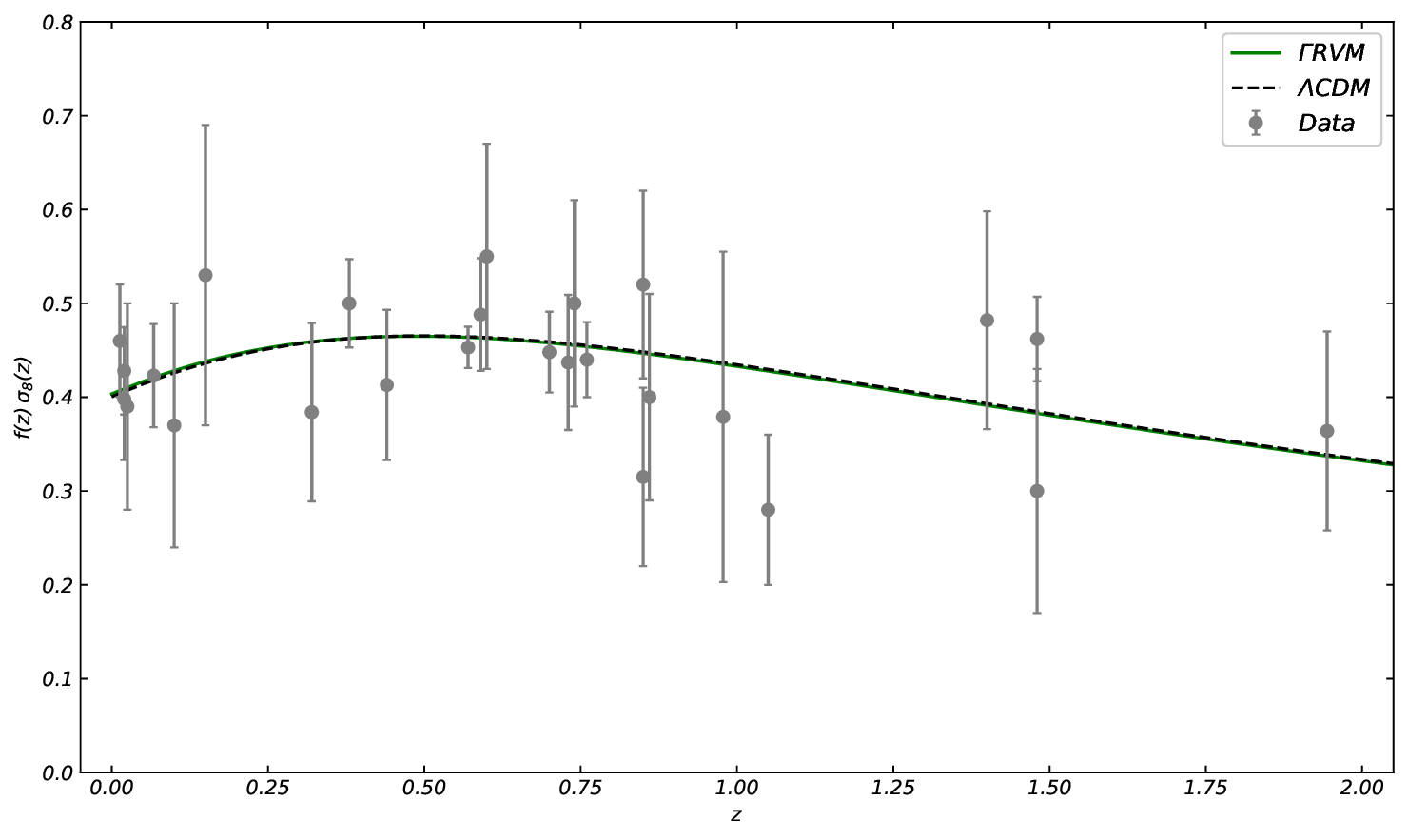}}
		\caption{Trajectories for $f(z)\sigma_8(z)$, shown for the standard model (black dashed) and the $\Gamma$RVM model (green). The curves are derived using the parameter constraints obtained from the BASE+$H_0$ dataset, given in Table~\ref{table1}.}
		\label{fig15}
	\end{minipage}
	\begin{minipage}[b]{0.85\textwidth}
		%\centering
		\scalebox{0.38}{\includegraphics{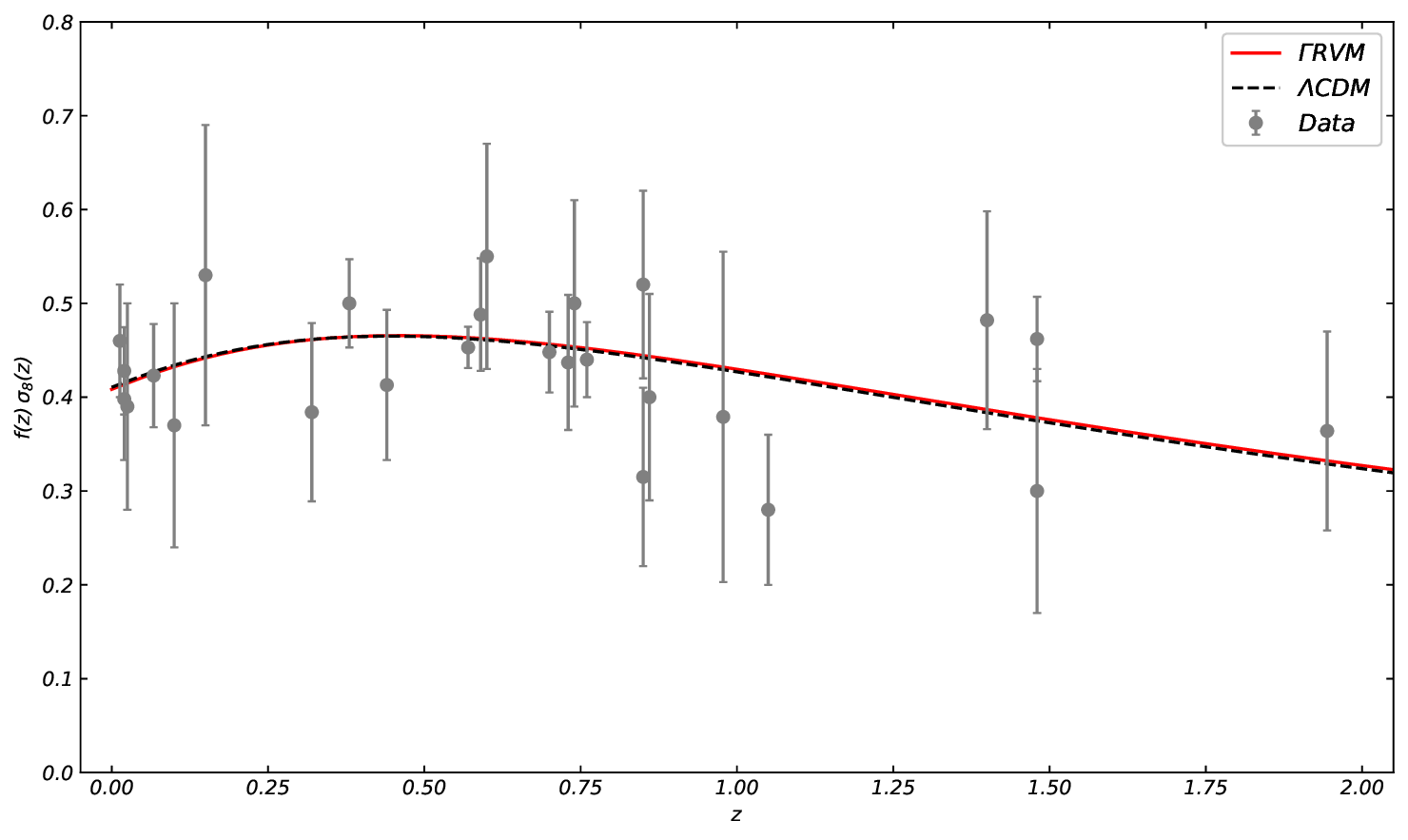}}
		\caption{Trajectories for the $f(z)\sigma_8(z)$, shown for the standard model (black dashed) and the $\Gamma$RVM model (red). The curves are derived using the parameter constraints obtained from the BASE+CMB+$H_0$ dataset, given in Table~\ref{table2}.}
		\label{fig16}
	\end{minipage}
\end{figure}

\section{Results and Discussion}
\label{sec6}
\begin{figure}[htbp]
\centering
	\begin{minipage}[b]{0.85\textwidth}
		%\centering
		\scalebox{0.45}{\includegraphics{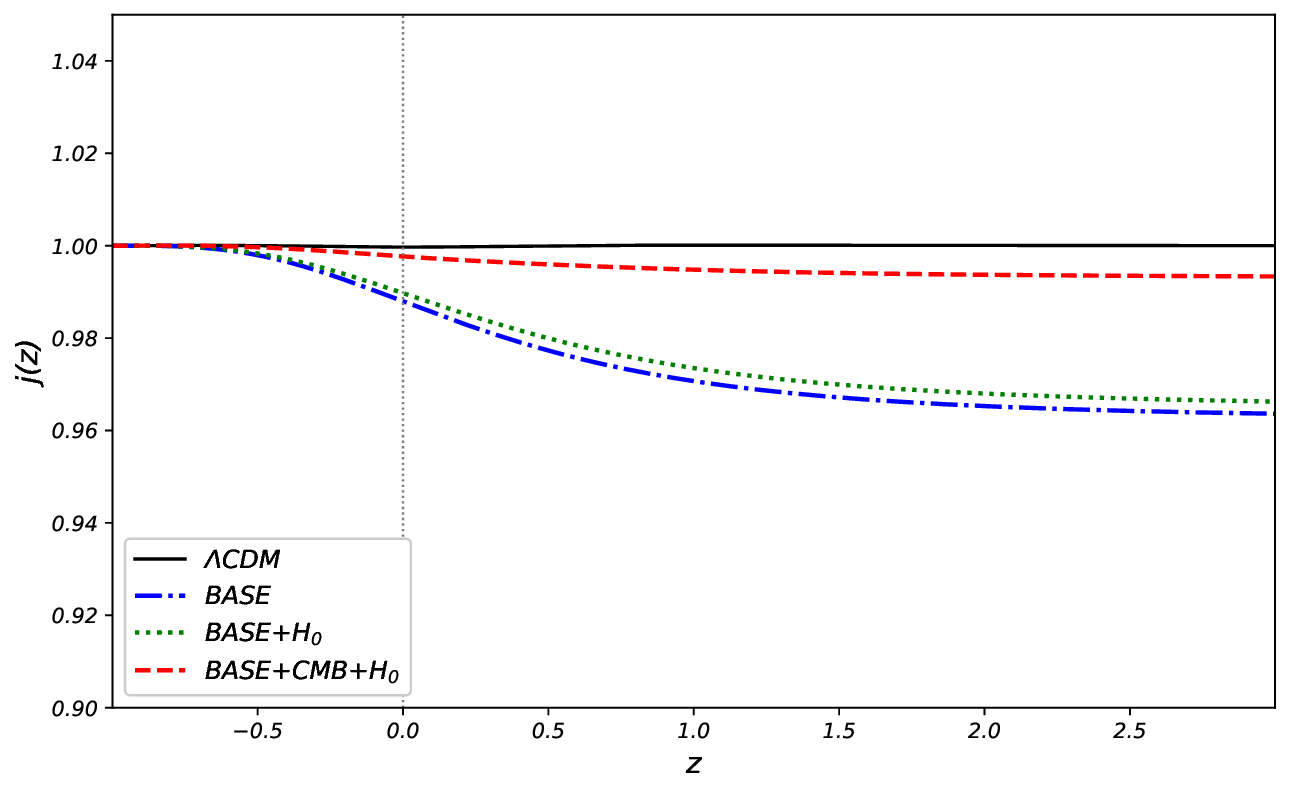}}
		\caption{Plot of the jerk parameter $j$ against redshift $z$ corresponding to the standard model (black line) for which it takes the value $1$ and the $\Gamma$RVM model from the parameter constraints obtained from the three datasets summarized in Table~\ref{table1} and Table~\ref{table2}. The parameter approaches 1 in the far future for all the cases.}
		\label{fig17}
	\end{minipage}
\end{figure}

\indent We now present the constraints on our model parameters based on the Hubble function, as defined in Eq.~\eqref{HubbleParameter}, using the observational datasets and the methodology described in the previous section. For the parameters of the $\Lambda$CDM and the $\Gamma$RVM models, Figs.~\ref{fig1}-\ref{fig3} display the combined $1-\sigma(68.3\%)$ and $2-\sigma(95.4\%)$ confidence level (CL) contours with marginalized likelihood distributions. Our chains are well converged as the values of the Gelman-Rubin statistic satisfy $\hat{R}<1.001$ for all the parameters across the three datasets. In Fig.~\ref{fig4} we display the combined contours and likelihood plots for the $\Gamma$RVM model across all three datasets. Clearly, the addition of the CMB distance priors provides tighter constraints.\\
\indent Table~\ref{table1} displays the matching parameter constraints acquired from the MCMC sampler for BASE and BASE+$H_0$, respectively, whereas Table~\ref{table2} reports the constraints for the data, including CMB distance priors. The constraints for the Hubble constant for the $\Gamma$RVM model are found to be $H_0 = 68.67^{+0.87}_{-0.86}$ km/s/Mpc for the BASE dataset, $H_0 = 70.53^{+0.67}_{-0.68}$ km/s/Mpc with the inclusion of the $H_0$ prior, and $H_0 = 68.33 \pm 0.40$ km/s/Mpc for the complete dataset including CMB distance priors. It is well known that the discrepancies in the measurement of $H_0$ between the Planck Collaboration~\cite{Planck18} and local cosmological probes~\cite{Riess22} constitute the so-called Hubble tension. The Planck collaboration reports $H_0 = 67.4 \pm 0.5$ km/s/Mpc, while R22 estimates a higher value of $H_0 = 73.04 \pm 1.04$ km/s/Mpc, leading to a tension at the level of $4.89\sigma$.
For the BASE dataset, the $H_0$ value is slightly higher than the Planck 2018 result, indicating that the late-time observations aren't sufficient to alleviate this tension. With the inclusion of the R22 prior on $H_0$, the estimated value shifts upward, showing a tension of $3.7\sigma$ with Planck and $2.0\sigma$ with R22. With the addition of the CMB distance priors, the inferred $H_0$ decreases relative to the BASE dataset, despite the $H_0$ prior, as expected since the distance priors are derived from Planck data and thus strongly favor the Planck value. While the $\Gamma$RVM model introduces modifications to the standard $\Lambda$CDM framework, it alone may not be sufficient to resolve the tension without further data refinements.
We plot the evolution of the Hubble parameter $H(z)$ for $\Lambda$CDM and $\Gamma$RVM model against the 32 data points of Cosmic Chronometers as shown in Figs.~\ref{fig5}-\ref{fig7}. Both models agree across the three datasets for the entire evolution period in late-time.\\
\indent In Figs.~\ref{fig8}-\ref{fig10}, we present the evolution of the deceleration parameter $q$ for the two models with respect to redshift $z$. There is a clear transition from a positive value of $q$ to a negative value, i.e. from a decelerating to an accelerating expansion phase, and we have derived the constraints on the transition redshift $z_{\text{tr}}$ for each dataset, displayed in Table~\ref{table1} and Table~\ref{table2}. For the BASE dataset, we have $z_{\text{tr}}=0.663 \pm 0.041 $, and with the $H_0$ prior, the values shifts earlier to $z_{\text{tr}}= 0.739^{+0.034}_{-0.035}$. Inclusion of the CMB data brings the value closer to $0.7$ as $z_{\text{tr}}= 0.697 \pm 0.015 $. There isn't a considerable difference with the values for the $\Lambda$CDM except for in the CMB case. We also report the present value of the deceleration parameter: $q_0= -0.537^{+0.024}_{-0.023}$, $-0.579^{+0.019}_{-0.018}$ and $-0.563 \pm 0.008$ for the three datasets, respectively. Comparing with the values of $q_0$ obtained for the $\Lambda$CDM model, we see that the current acceleration is slower in the case of BASE and BASE+$H_0$ for our proposed model, whereas with CMB, the Universe is accelerating at a slightly faster rate. This is, of course, expected due to the influence of the early-time physics introduced by the distance priors.\\
\indent We now turn to the model-specific parameters of the $\Gamma$RVM model: the vacuum decay parameter $\nu$ and the matter creation parameter $\beta$. Theoretically, $\nu$ is expected to be close to zero, of order $10^{-3}$, while $\beta$ must satisfy $\beta > 0$. Our constraints are $\nu = 0.00218^{+0.00141}_{-0.00124}$ (BASE), $\nu = 0.00209^{+0.00138}_{-0.00122}$ (BASE+$H_0$), and $\nu = 0.00067^{+0.00036}_{-0.00034}$ (BASE+CMB+$H_0$). For the matter creation parameter $\beta$, we obtain $\beta = 0.00628^{+0.00378}_{-0.00349}$ (BASE), $\beta = 0.00579^{+0.00363}_{-0.00333}$ (BASE+$H_0$), and $\beta = 0.00087^{+0.00045}_{-0.00042}$ (BASE+CMB+$H_0$). The inclusion of CMB data strongly suppresses both parameters, driving them closer to zero due to tensions with early-time physics. Despite being of order $10^{-4}$, the values are still able to impact the Universe’s expansion history. We can see this in the estimate for $r_d$, which gives the value of the sound horizon at drag epoch. The Planck estimate for this is around $147$ Mpc, and with our data, we obtain $r_d = 147.35^{+0.23}_{-0.22}$ Mpc for $\Lambda$CDM, which is consistent. For the proposed model, the obtained value is higher, with $r_d = 149.96^{+0.89}_{-0.84}$ Mpc, which has a moderate tension of around 3$\sigma$ with Planck 2018 TT,TE,EE+lowE estimate. Since $r_d$ sets the calibration of the acoustic scale, this tension has direct implications for both CMB and BAO, and thus the $\Gamma$RVM model clearly deviates from the $\Lambda$CDM expectations in early times. \\
\indent  In the considered cosmological scenario, the values of the age of the Universe $t_0$ based on the parameter constraints obtained for the three datasets are $13.70 \pm 0.09$ Gyr, $13.69 \pm 0.09$ Gyr, and $13.92 \pm 0.06$ Gyr, respectively. The values are comparable for the first two datasets with the standard model estimates; however, CMB suggests a longer cosmic history. The age estimates for $\Lambda$CDM are $13.67 \pm 0.09$ Gyr, $13.66 \pm 0.09$ Gyr, and $13.75 \pm 0.02$ Gyr, and we can clearly see the results are consistent for BASE and BASE+$H_0$. However, for our third dataset, the difference in $t_0$ is about $2.7\sigma$, which is statistically significant. This can be attributed to the decaying vacuum and matter creation dynamics whose impact on the Universe's evolution is considerable despite their respective parameters $\nu$ and $\beta$ being close to zero.\\
\indent In this study, we have assumed the EoS of the running vacuum as $w_{\Lambda}=-1$. To assess the overall EoS, we investigate the effective EoS given in \eqref{eff_EoSpara}. The evolution of EoS is depicted in Figs.~\ref{fig11}-\ref{fig13}. From the figures, we clearly see that $w_{\text{eff}}$ has a negative value such that $w_{\text{eff}} <  -1/3 $, which indicates an accelerating phase in the expansion. In the future, we can see that the model asymptotically approaches negative unity, i.e. $w_{\text{eff}}=-1$, implying complete vacuum domination. The present values of $w_{\text{eff}}$, represented by $w_0^{\text{eff}}$ are $-0.691^{+0.016}_{-0.015}, -0.719 \pm 0.012, -0.708 \pm 0.005$ for the $\Gamma$RVM model. The values are consistent with those reported for the $\Lambda$CDM model (Tables ~\ref{table1} and ~\ref{table2}).\\
\indent Let us now discuss $\sigma_8$ and $S_8$, which are crucial for structure formation and are obtained by probing cosmological perturbations, as discussed in Section~\ref{sec4}. For the proposed model, the $\sigma_8$ constraints are $0.802^{+0.024}_{-0.026}, 0.818 \pm 0.023,$ and $0.809 \pm 0.022$ respectively, for the three datasets. We are particularly interested in assessing the parameter $S_8$ which is derived from $\sigma_8$ and the matter density parameter $\Omega_m$ through the relation: $S_8 = \sigma_8 (\Omega_m/0.3)^{0.5}$. This parameter is associated with the growth of cosmic structures and quantifies the clustering of matter in different regions of space. The Planck collaboration reports a value of $S_8 = 0.831 \pm 0.013$, while a number of recent weak-lensing surveys~\cite{Busch22,Amon22,Dalal23} have instead favored a Universe less rich in structure, with values below $0.78$. In our analysis, we find $S_8 = 0.816 \pm 0.026$ for the \textsc{BASE} dataset, which is fully consistent with the Planck estimate. When extending to the \textsc{BASE}+$H_0$ dataset, we obtain $S_8 = 0.794^{+0.023}_{-0.024}$, still well within the Planck uncertainties. Including CMB data yields a nearly identical estimate, $S_8 = 0.797^{+0.022}_{-0.021}$. Furthermore, we find no statistically significant difference between the constraints derived within the standard cosmological framework and those obtained under the proposed extension.
For comparison, the joint KiDS-1000 and DES Y3 analysis~\cite{Abb23} reported $S_8 = 0.790^{+0.018}_{-0.014}$, while the more recent KiDS-Legacy cosmic shear combined-probes analysis~\cite{Wright25} obtained $S_8 = 0.815^{+0.016}_{-0.021}$, a value in excellent agreement with Planck. Taken together, our results are consistent with these latest measurements and do not point to any significant deviation from the $\Lambda$CDM prediction.\\
\indent We present the plots for the weighted linear growth rate function $f(z)\sigma_8(z)$ for the three datasets using the parameter estimates corresponding to both the standard model and the proposed model in Figs.~\ref{fig14}-\ref{fig16}. We do the comparison against the 26 data points of the $f(z)\sigma_8(z)$ measurements that have been used. The resulting curves provide a fairly good fit for the data, implying our model agrees with the observational data in the context of structure formation and cosmic perturbations. \\
\indent Furthermore, we report the cosmographic parameter, known as the jerk parameter, which is defined by $j = \dddot{a}/aH$. It gives information about the dynamics of the DE EoS. In terms of the deceleration parameter, it is given by the expression:
\begin{equation}
	j = q(2q + 1) + (1 + z) \frac{dq}{dz}.
	\label{jerk_para}
\end{equation}
It can be verified that for the standard model, the value of the jerk parameter is a constant, i.e. $j=1$. This parameter is a dimensionless third derivative of the scale factor and can help us study how our model deviates from the standard model. For the proposed model, solving the Eq.~\eqref{jerk_para} gives
\begin{equation}
	j = 3(1-\beta)(1-\nu)(q+1)-3q-2.
	\label{jerk_para_model}
\end{equation}
We see that in the limit where the model parameters $\beta$ and $\nu$ are zero, $j$ reduces to one. In the case of the deceleration parameter, negative values imply an accelerating expansion of the Universe. However, in the case of the jerk parameter, a positive value represents the same, and from \eqref{jerk_para_model}, we can see that for the considered model, the value will remain positive. In Fig.~\ref{fig17}, the evolution of this parameter is shown for our model using the parameter constraints obtained for all three datasets. The figure confirms the positive nature, and we see that it remains close to but less than unity at early redshifts, eventually tending to 1 in the future. The current values of the jerk parameter $j_0$ are estimated to be around 0.988, 0.990 and 0.998 for the three datasets. Thus, the parameter deviates only slightly from the standard model, and in the future, it converges to a similar behaviour. \\

\section{Selection Criteria}
\label{sec7}
\noindent In this section, we are interested in determining the model that performs better with respect to the observational data used in the analysis. We utilize three selection statistical criteria, namely, reduced chi-squared test, Akaike information criteria (AIC) and Deviance information criteria (DIC).\\
\indent Table~\ref{table3} presents the $\chi^2$ and the reduced $\chi^2$ values for the standard model and the proposed model, respectively. The reduced $\chi^2$, denoted by $\chi^2_{red}$, is calculated by the formula ${\chi^2}_{red} = \chi^2/(N-d)$, where $N$ is the total number of data points, and $d$ is the number of free parameters. It is to be noted that a reduced $\chi^2$ value less than one gives a good fit, and greater than one indicates a poor fit. From Table~\ref{table3}, we see that the reduced $\chi^2$ values are less than 1, indicating that our model fits well with the observational datasets and the data fits well with the current models. This is just one criterion to test the viability of our model. We will utilize two selection criteria for further analysis.\\

\indent The Akaike information criteria (AIC)~\cite{Aka74} value is calculated as
\begin{equation}
	AIC = \chi^2_{min} + \frac{2d\;N}{N-d-1},
	\label{AIC}
\end{equation}
where $\chi^2_{min}$ is the minimum value of the $\chi^2$-function. Eq.~\eqref{AIC} makes it clear that the penalty for the extra number of free parameters is represented by the second term on the right-hand side. In cases where $d$ is significantly lesser than $N$, Eq.~\eqref{AIC} reduces to $AIC = \chi^2_{min} + 2d$. To test the proposed model $M_1$ against the reference model $M_0$, we calculate the difference in their respective AIC values as $\Delta {AIC}_{M_1 M_0} = {AIC}_{M_1} - {AIC}_{M_0} $, which is interpreted as the "evidence in favour" of the model $M_1$ as compared to the reference $M_0$. In this paper, the $\Lambda$CDM model has been considered as the reference.\\
\indent The value $\Delta {AIC}_{M_1 M_0} < 2$ indicates "strong evidence in favour" of the proposed model $M_1$. For $ 2 \leq \Delta {AIC}_{M_1 M_0} \leq 4 $, there is "averagely strong evidence in favour" of $M_1$. If   $ 4 < \Delta {AIC}_{M_1 M_0} \leq 7 $, then there is "little evidence in favour", whereas for $\Delta {AIC}_{M_1 M_0} > 7 $, there is "no evidence in favour" of the proposed model.\\

\indent The Deviance Information Criterion (DIC)~\cite{Spieg02} is a Bayesian model selection tool that incorporates both model fit and complexity, using the full posterior distribution rather than relying only on the best-fit values that minimize the chi-squared or maximize the likelihood. It is computed from the entire Markov Chain Monte Carlo (MCMC) sample and is given by:
\begin{equation}
  \text{DIC} = 2\tilde{D} - D(\langle \theta \rangle),
  \label{dic}
\end{equation}
where $\tilde{D} = \langle D(\theta) \rangle = \langle -2 \ln \mathcal{L}(\theta) \rangle$ is the mean deviance across the posterior, and $D(\langle \theta \rangle) = -2 \ln \mathcal{L}(\langle \theta \rangle)$ is the deviance evaluated at the mean of the posterior parameter values. Models with lower $DIC$ values are favored over those with higher values. In other words, the greater the difference between the DIC values of model $M_1$ and reference $M_0$, the less favored will be the model $M_1$.\\

\begin{table*}[htbp]
	\centering
	\begin{tabular}{|c|c|c|c|c|c|c|}
		\hline
		\textbf{Values} & \multicolumn{2}{c|}{\textbf{BASE}} & \multicolumn{2}{c|}{\textbf{+$H_0$}} & \multicolumn{2}{c|}{\textbf{+CMB}} \\
\cline{2-7}
		& \textbf{$\Lambda$CDM} & \textbf{$\Gamma$RVM} & \textbf{$\Lambda$CDM} & \textbf{$\Gamma$RVM} & \textbf{$\Lambda$CDM} & \textbf{$\Gamma$RVM} \\ \hline
		$\chi^2$        & 1080.72               & 1081.36       & 1092.80              & 1088.08       & 1094.59              & 1082.82   \\
		$d$             & 3                     & 5              & 3                     & 5              & 4                     & 6          \\
		$N$             & 1118                 & 1118           & 1118                  & 1118          & 1121                  & 1121       \\
		$\chi^2_{\rm red}$ & 0.97                & 0.97          & 0.98                 & 0.98          & 0.98                & 0.97      \\
		AIC             & 1086.74               & 1091.42        & 1098.82               & 1098.14       & 1102.63              & 1094.90    \\
        $\Delta$AIC     & ---                   & 4.68          & ---                   & -0.68         & ---                   & -7.73      \\
		DIC            & 1087.01               & 1087.30        & 1091.41               & 1091.42        & 1094.04               & 1092.24    \\
		$\Delta$DIC     & ---                   & 0.29         & ---                   & 0.01         & ---                   & -1.80      \\ \hline
	\end{tabular}
    \caption{Comparison of $\chi^2$, AIC, and DIC values for $\Lambda$CDM and the $\Gamma$RVM model across the three datasets.}
	\label{table3}
\end{table*}

\indent The values of $\Delta \mathrm{AIC}$ and $\Delta \mathrm{DIC}$ are shown in Table~\ref{table3}. For the three datasets, the $\Delta \mathrm{AIC}_{M_1 M_0}$ values are 4.68, -0.68, and -7.73, respectively. For the BASE dataset, the value is between 4 and 7, indicating little evidence in favour of the proposed model and thus, in this case, $\Lambda$CDM is preferred. With the inclusion of the $H_0$ prior, the evidence shifts in favor of our model as the value becomes slightly less than zero. Furthermore, when the CMB distance priors are included, there is a stronger preference for the proposed $\Gamma$RVM model as the negative magnitude is higher.\\
\indent When comparing DIC, there is no clear preference between the models for the BASE and BASE+$H_0$ datasets, with $\Delta \mathrm{DIC}$ values of 0.29 and 0.01, respectively. These values are close to zero and thus not indicative of any significant evidence. In the case of the complete dataset where CMB distance priors are included, the $\Delta \mathrm{DIC}$ value is -1.80, indicating a slight preference for the $\Gamma$RVM model over the standard model.
These results show a trend where the proposed model gains more support as the dataset becomes increasingly comprehensive. Although the full CMB spectrum is not considered, the inclusion of distance priors is able to provide a clear picture that the additional parameters in $\Gamma$RVM improve the overall model fit. This suggests that the robust modelling approach may offer a viable alternative to $\Lambda$CDM with more refined data.

\section{Thermodynamic Analysis}
\label{sec8}
In this section, we focus on the thermodynamic analysis of a Universe under the proposed scenario of vacuum decay into matter. First, we use the generalized first law of thermodynamics to obtain an expression for the dark matter temperature in terms of our model parameters. Then, we test the validity of the generalized second law (GSL) of thermodynamics by computing the total entropy of the Universe and analyzing its evolution with respect to cosmic time. We also incorporate the Casimir effect into our framework and assess how it affects the thermodynamic properties.\\
\noindent We note that thermodynamic aspects of running vacuum models (RVMs), particularly in the context of particle and entropy production and the GSL, have been studied in detail in previous works~\cite{SolaYu20}. These considerations closely relate to the present analysis and support the physical consistency of matter creation models with vacuum decay.

\subsection{Dark Matter Temperature}
\label{sec8.1}
For computing the temperature of dark matter in the proposed scenario, we take into consideration the work done by the authors in the studies~\cite{Harko13,Iva19}. First, we consider the modified first law of thermodynamics, given by Eq.~\eqref{eq:first_law_modified}. This can be rewritten as
\begin{equation}
	d(\rho_m V) + \tilde{p}_m \, dV - \frac{(\rho_m+\tilde{p}_m)}{n} d(nV) = 0, \label{t1}
\end{equation}
where $h=(\rho_m+\tilde{p}_m)$. Taking time derivative of Eq.~\eqref{t1}, we have
\begin{equation}\label{t2}
\dot{\rho_m}=\frac{\dot{n}}{n}(\rho_m+\tilde{p}_m).
\end{equation}
Since the Gibbs equation defines the temperature, we have
\begin{equation}\label{t3}
TdS=d\left(\frac{\rho_m}{n}\right)+\tilde{p}_m\;d\left(\frac{1}{n}\right)
\end{equation}
We have $T=T(n, \rho_m)$ which gives
\begin{equation}\label{t4}
\dot{T}=\frac{\partial T}{\partial n}\dot{n}+\frac{\partial T}{\partial \rho_m}\dot{\rho_m}.
\end{equation}
Using \eqref{t2} into \eqref{t4}, we get
\begin{equation}\label{t5}
\dot{T}=\frac{\dot{n}}{n}\left[\frac{\partial T}{\partial n}n+\frac{\partial T}{\partial \rho_m}(\rho_m+\tilde{p}_m)\right].
\end{equation}
Using the integrability condition, $\partial^2 S/\partial T \partial n=\partial^2 S/\partial n \partial T $, one gets
\begin{equation}\label{t6}
\frac{\partial T}{\partial n}n+\frac{\partial T}{\partial \rho_m}(\rho_m+\tilde{p}_m)=T\frac{\partial \tilde{p}_m}{\rho_m}.
\end{equation}
Using \eqref{t6} into \eqref{t5}, we obtain the evolution equation for the temperature, which is given by
\begin{equation}\label{t7}
\frac{\dot{T}}{T}=\frac{\dot{n}}{n}\frac{\partial \tilde{p}_m}{\partial \rho_m}=w_{\text{eff}}\frac{\dot{n}}{n}.
\end{equation}
Using Eq.\eqref{balance} in the above equation, we get
\begin{equation}\label{t8}
\frac{\dot{T}}{T}= -3w_{\text{eff}}H\left(1-\frac{\Gamma}{3H}\right),
\end{equation}
or equivalently,
\begin{equation}\label{t9}
\frac{\dot{T}}{T}=3\beta (1-\beta)H.
\end{equation}
Using \eqref{scale} into \eqref{t9}, we obtain the temperature of dark matter
\begin{equation}
T=T_0\left[\sqrt{\frac{\tilde{\Omega}_m}{\tilde{\Omega}_{\Lambda}}}\sinh\left(\frac{3}{2}\sqrt{\tilde{\Omega}_{\Lambda}}(1-\beta)(1-\nu)H_0t\right)\right]^{\frac{2\beta}{(1-\nu)}}.
\end{equation}
Using \eqref{scale} into \eqref{balance}, we can obtain the particle number density, given by
\begin{equation}
n(t)=n(t_0)\left[\sqrt{\frac{\tilde{\Omega}_m}{\tilde{\Omega}_{\Lambda}}}\sinh\left(\frac{3}{2}\sqrt{\tilde{\Omega}_{\Lambda}}(1-\beta)(1-\nu)H_0t\right)\right]^{-\frac{2}{(1-\nu)}}.
\end{equation}

\begin{figure}[htbp]
\centering
	\begin{minipage}[b]{0.85\textwidth}
		%\centering
		\scalebox{0.4}{\includegraphics{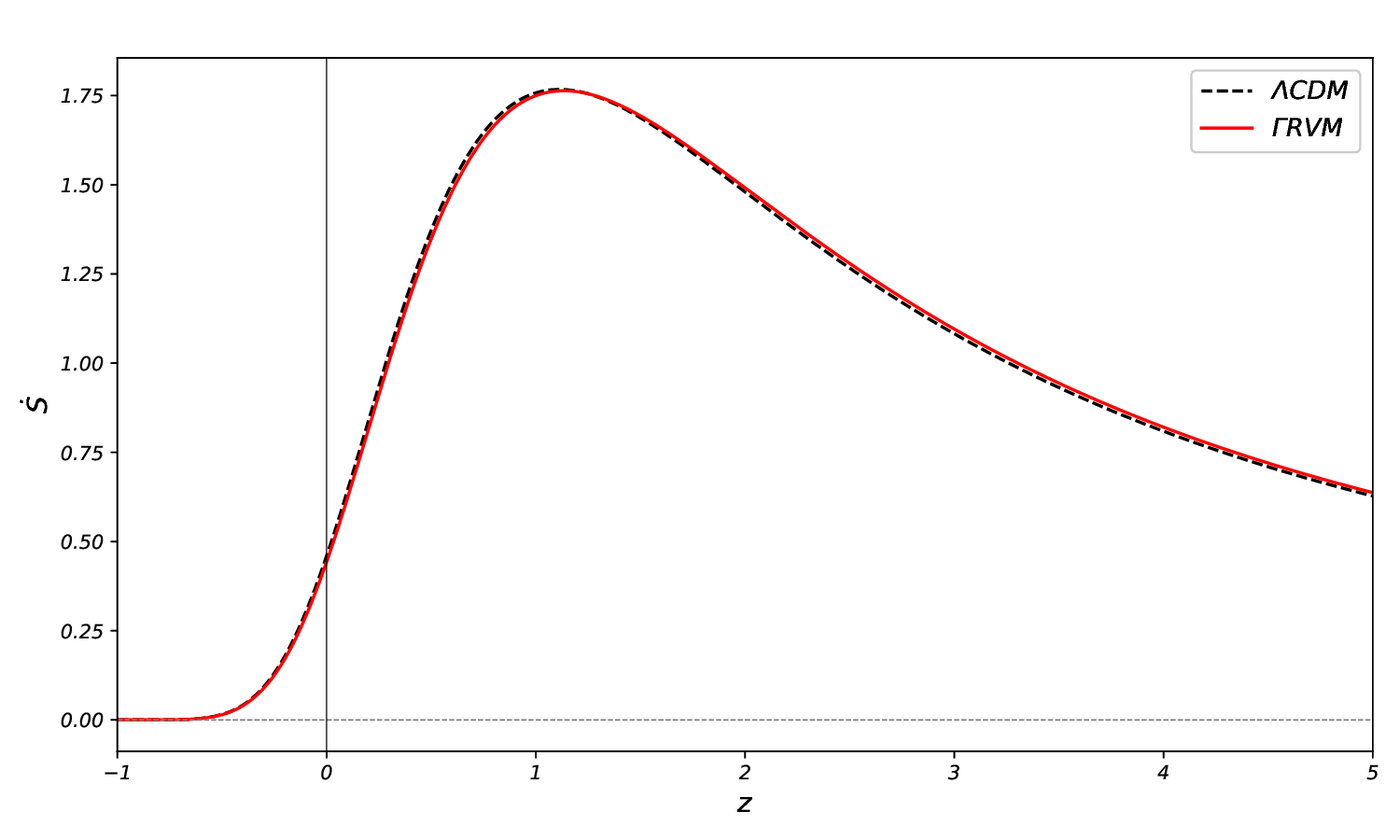}}
		\caption{Plot of the rate of change of entropy with respect to redshift $z$ obtained from the parameter constraints for the BASE + CMB + $H_0$ data. The positive value ensures that the entropy always increases. The dashed line shows the evolution for the standard model and red line for the $\Gamma$RVM model.}
		\label{fig18}
	\end{minipage}
	\begin{minipage}[b]{0.85\textwidth}
		%\centering
		\scalebox{0.4}{\includegraphics{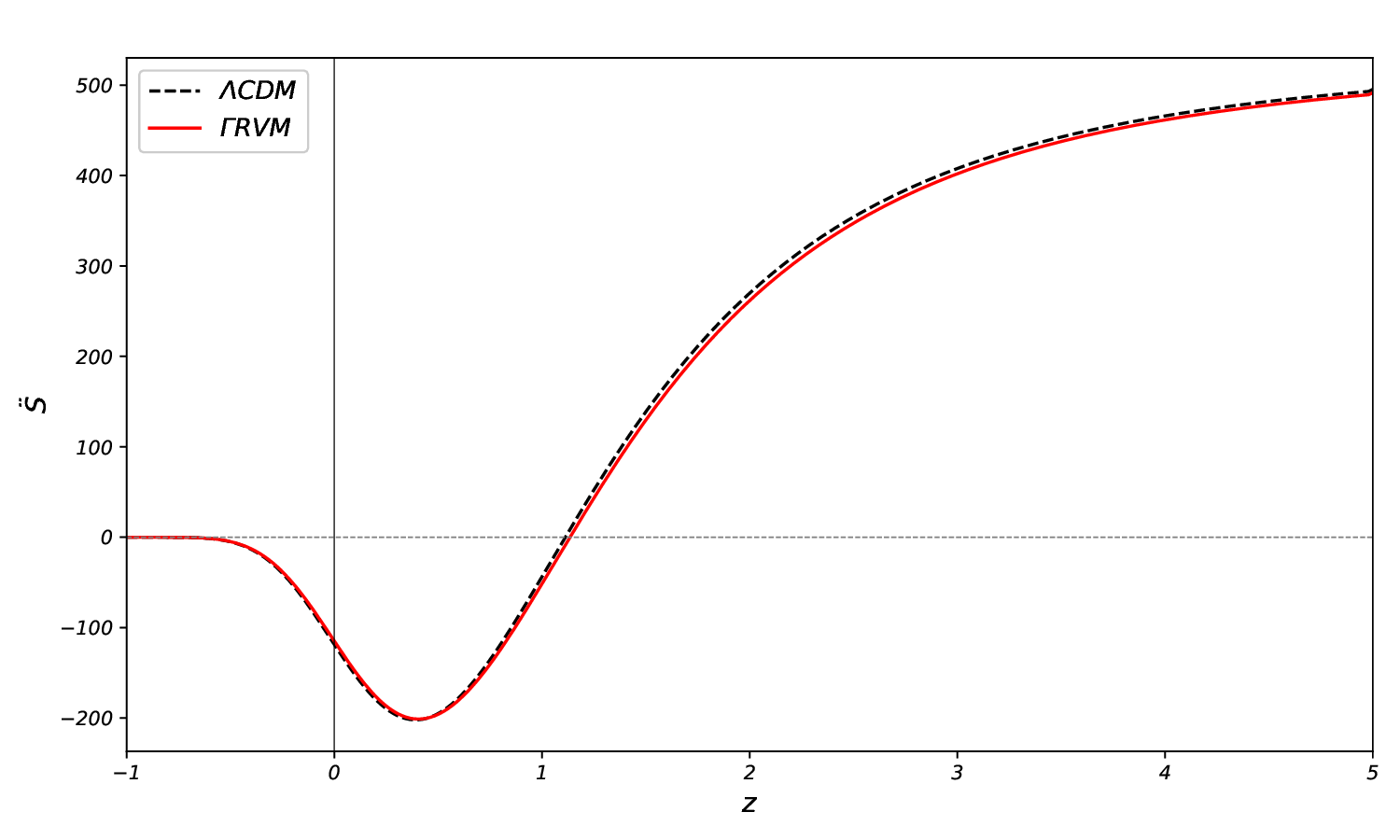}}
		\caption{Plot of the acceleration of entropy with respect to redshift $z$ obtained from the parameter constraints for the BASE + CMB + $H_0$ data. The dashed line shows the evolution for the standard model and red line for the $\Gamma$RVM model.}
		\label{fig19}
	\end{minipage}
\end{figure}

\subsection{Generalized Second Law of Thermodynamics}
\label{sec8.2}
This section is devoted to examining and analyzing the validity of the generalized second law (GSL) of thermodynamics for the presented model. According to the GSL, the total entropy of the Universe, i.e., the sum of the horizon entropy and entropy of the fluid, always increases with time. Mathematically, we can express it as
\begin{equation}\label{GSL}
  \frac{d}{dt} (S_H+S_m+S_{\Lambda})\geq 0,
\end{equation}
where $S_H$ is the entropy associated with the Hubble horizon (apparent horizon) for a flat Universe, $S_m$ is the entropy of dark matter produced by gravitationally induced matter creation, and $S_{\Lambda}$ is the entropy associated with vacuum energy. The GSL thermodynamics was first formulated for black holes by Bekenstein and Hawking~\cite{Bek73}. They stated that the black hole horizons have entropy and temperature associated with them. According to Bekenstein, the horizon entropy is proportional to the area of the horizon, which is given by
\begin{equation}\label{horizon_entropy}
  S_H = \frac{k_B A}{4 l_{\text{Pl}}^2},
\end{equation}
where $k_B$ is the Boltzmann constant, $A=4 \pi {R_H}^2$ is the area of the Hubble horizon and $l_{\text{Pl}}=\sqrt{\frac{\hbar G}{c^3}}$ is the Planck length. Here, $R_H = \frac{1}{\sqrt{(H^2+\kappa a^{-2})}}$ is the radius of the Hubble horizon. In a spatially flat FLRW Universe, we have $R_H=\frac{1}{H}$. Assuming $k_B=\hbar=c=8\pi G=1$, the above equation can be rewritten as
\begin{equation}\label{S_horizon}
  S_H = \frac{8 \pi^2}{H^2}
\end{equation}
Therefore, the rate of change of entropy of the Hubble horizon is given by
\begin{equation}\label{SH_dot}
  \dot{S}_H=-\frac{16 \pi^2}{H^3} \dot{H}
\end{equation}
The entropy of cosmic components (matter and dark energy) inside the Hubble horizon can be calculated by the Gibbs equation~\cite{Pav06}:
\begin{equation}\label{Gibbs_eqn}
  T \, dS = d(\rho V) + p \, dV,
\end{equation}
where $V=4 \pi / 3 H^3$ is volume enclosed by horizon and $T$ is temperature associated with both matter and dark energy and due to their mutual interaction, we assume $T$ to be equal across the dark fluid. We assume that the system bounded by the Hubble horizon remains in equilibrium so that the temperature distribution is uniform and its value is equal to the temperature of the horizon. In this case, the Gibbons-Hawking temperature, $T=H/2 \pi$, is a natural choice for the horizon temperature.\\
\indent It is evident from \eqref{Gibbs_eqn} that the vacuum energy density in the present model doesn't contribute to the entropy as $\rho_{\Lambda}=-p_{\Lambda}$. The non-zero contribution to the entropy is only due to dark matter produced by gravitationally induced matter creation, which is given by
\begin{equation}\label{matter_entropy}
  S=\frac{(\rho_m+\tilde{p}_m) V}{T},
\end{equation}
where $\tilde{p}_m=p_m+p_c$. Substituting $V=4 \pi / 3 H^3$ and $T = H/2 \pi$ in above equation, we get matter entropy as
\begin{equation}\label{S_matter}
  S_m = \frac{8 \pi^2}{3 H^4} (\rho_m+p_c).
\end{equation}
Using $p_c=-\beta \rho_m$, the above equation can be written as
\begin{equation}\label{S_matter_exp}
  S_m = \frac{8 \pi^2 (1 - \beta)\rho_m}{3 H^4},
\end{equation}
where matter energy density $\rho_m$ is obtained from the Friedmann equation \eqref{eq:FE1}, which is given by
\begin{equation}\label{matter_energy_density_rho_m}
  \rho_m = 3 H_0^2 (1 - \nu) \tilde{\Omega}_m a^{-3(1-\beta)(1-\nu)}.
\end{equation}
Using \eqref{S_horizon} and \eqref{S_matter_exp} the total entropy of the Universe, i.e., $S=S_H+S_m$ is calculated as
\begin{equation}\label{entropy}
  S = \frac{8 \pi^2}{H^2} \left[1 + \frac{(1-\beta)\rho_m}{3 H^2}\right].
\end{equation}
On substituting the value of $\rho_m$ from \eqref{matter_energy_density_rho_m}, we see that the total entropy can be simplified in terms of deceleration parameter $q$, given in \eqref{dec_para}, as
\begin{equation}\label{S_total}
  S = \frac{8 \pi^2}{3 H^2} (5 + 2q).
\end{equation}
Now, to check the validity of GSL, given by \eqref{GSL} for this model, we have the combined change in entropy for the dark fluid
\begin{equation}\label{Sdot_darkfluid}
  \dot{S}_m = -\frac{8 \pi^2 (1 - \beta) \rho_m}{H^3} \left(1 + \frac{\dot{H}}{H^2}\right).
\end{equation}
Combining \eqref{SH_dot} and \eqref{Sdot_darkfluid}, and noting Eq. \eqref{eq:deceleration_parameter}, we obtain the total change in entropy as
\begin{equation}\label{entropy_change}
  \dot{S}=\frac{8 \pi^2}{H}\left[2(1+q)+\frac{(1-\beta)\rho_m}{H^2}q\right],
\end{equation}
in which we substitute the value of $\rho_m$ and solve further to obtain the simplified expression
\begin{equation}\label{Sdot_total}
  \dot{S}=\frac{16 \pi^2}{H}{(1+q)}^2.
\end{equation}
In equation \eqref{Sdot_total}, it is noted that $\dot{S} \geq 0$ as all the terms are positive and hence the GSL is satisfied. In the asymptotic limit, the change in entropy tends to zero in late time, which shows that the entropy is extremized and thus the end phase is in an equilibrium state. As $q \to -1$ in the asymptotic future, the total entropy $S$ becomes equal to the horizon entropy $S_H$, given by
\begin{equation}\label{entropy_future}
  S_{\text{max}} = \frac{8 \pi^2}{H_0^2 \tilde{\Omega_{\Lambda}}}
\end{equation}
The change in entropy, $\dot{S} \to 0$ implies that the total entropy stabilizes at the value $S_{\text{max}}$ and does not increase any further, i.e. it reaches an asymptotic limit ~\cite{Fraut82}. \\
\indent The evolution of $\dot{S}$ with respect to redshift $z$ is shown in Fig.~\ref{fig18} using the parameter constraints obtained for BASE+CMB+$H_0$ data. The figure shows that the time derivative of entropy, $\dot{S}$ remains positive throughout the cosmic evolution and tends towards zero in the future. We also plot the evolution of $\ddot{S}$ with respect to redshift, which is shown in Fig.~\ref{fig19}. It is observed that $\ddot{S}>0$ in the early phase of evolution and made a transition to $\ddot{S}<0$ in the recent past. A thermodynamically stable equilibrium state is achieved as $z \rightarrow -1$; $\ddot{S}\rightarrow 0$, which shows that any system satisfying the extremum of entropy and convexity condition behaves like an ordinary macroscopic system. Therefore, we can say that the evolution of the Universe is like the evolution of an ordinary macroscopic system. Overall, the results are consistent with established thermodynamic principles.

\subsection{$\Gamma$RVM with Casimir effect}
\label{sec8.3}
\indent We now consider the Casimir effect, which in essence is a polarization of the vacuum of quantized fields. This phenomenon causes the spectrum of vacuum oscillations to change due to the bounded quantization volume or a space having a non-Euclidean topology. The effect has applications in numerous branches of physics, especially in cosmology, since it is a prime example where a non-Euclidean space is fundamental~\cite{Moste88}. \\
\indent The Casimir effect is relevant in our analysis due to the decaying vacuum (RVM) framework, which comes from quantum field theory. In QFT, the zeropoint fluctuations of the quantum fields contribute to the energy of the vacuum. However, this energy has not yet been observed in any laboratory experiment. The Casimir energy, which arises from the effect in question, is often considered a potential candidate for these fluctuations. We shall refrain from going into details of QFT physics of Casimir effect and refer the reader to~\cite{Jaffe05} for an adequate discussion on the matter.\\
\indent In the standard laboratory setting, the Casimir effect arises between two closely spaced conducting plates\cite{Casimir48}, whose presence imposes boundary conditions on the quantum fields. In a cosmological context, a common approach is to regard the apparent horizon as the relevant boundary; i.e., $R_H$ sets the scale for the Casimir energy. A key property of this energy is that the pressure inside a bounded region exceeds that outside its corresponding shell. In the following analysis, we follow a similar approach as adopted by the authors in~\cite{Setare10,Akumar21}. Our primary concern is checking if the GSL of thermodynamics still holds with Casimir contributions.\\
\indent Motivated by the expectation that Casimir energy is most significant at early times and diminishes at late epochs, we adopt the following expression for the Casimir energy:
\begin{equation}\label{CasimirEnergy}
  E_{\mathrm{cs}} = \frac{\epsilon}{R_H},
\end{equation}
where $\epsilon$ is a proportionality constant. The inverse dependence on $R_H$ reflects the theoretical scaling predicted for a finite causal domain. We note that $\epsilon$ is phenomenological and depends on the underlying theory.\\
\indent For the given Casimir energy, the corresponding pressure can be obtained by calculating the force per unit area, which can be written as:
\begin{equation}\label{CasimirPressure}
  p_{\mathrm{cs}} = - \frac{1}{A} \frac{\partial E_{\mathrm{cs}}}{\partial R_H},
\end{equation}
where $A = 4 \pi {R_H}^2$ is the area. On substituting the value of $E_{\mathrm{cs}}$ from above, we obtain
\begin{equation}\label{p_cs}
  p_{\mathrm{cs}} = \frac{\epsilon}{4 \pi {R_H}^4}
\end{equation}
From this, we can see that the Casimir energy density $\rho_{\mathrm{cs}}$ evolves with the apparent horizon radius as $\rho_{\mathrm{cs}} \propto {R_H}^{-4} $. The continuity equation for the Casimir energy density can thus be written as:
\begin{equation}\label{casimir_continuity}
  \dot{\rho_{\mathrm{cs}}} + 3 H (1 + w_{\mathrm{cs}}) \rho_{\mathrm{cs}} = 0.
\end{equation}
Here, $w_{\mathrm{cs}}$ is the equation of state parameter for the Casimir energy. We can calculate the energy density using $\rho_{\mathrm{cs}} = E_{\mathrm{cs}} / V$ using the value of the Casimir energy from Eq.~\eqref{CasimirEnergy} and $ V = 4 \pi {R_H}^3 / 3$. It can be further deduced that the EoS parameter $w_{\mathrm{cs}} = 1/3$. This allows us to write the time derivative of the Casimir energy density as $\dot{\rho_{\mathrm{cs}}}=-4 H \rho_{\mathrm{cs}}$.

Since we have assumed $8 \pi G = 1$ throughout our thermodynamic analysis, the Friedmann equation \eqref{eq:FE1} with the Casimir energy in consideration can be written as:
\begin{equation}\label{fe_cs}
  3 H^2 = \rho_m + \rho_\Lambda + \rho_\mathrm{cs}.
\end{equation}
Taking the derivative with respect to time and using the continuity equation for the dark sector, we get:
\begin{equation}\label{fe_cs_diff}
  \dot{H} = - \left[\frac{(1-\beta)\rho_m}{2} + \frac{2}{3} \rho_\mathrm{cs} \right].
\end{equation}
Rearranging the terms, we get an expression for the Casimir energy density as
\begin{equation}\label{rho_cs}
    \rho_\mathrm{cs} = - \frac{3}{2} \left(\dot{H} + \frac{(1-\beta)\rho_m}{2} \right).
\end{equation}

\begin{figure}[htbp]
\centering
	\begin{minipage}[b]{0.85\textwidth}
		%\centering
		\scalebox{0.4}{\includegraphics{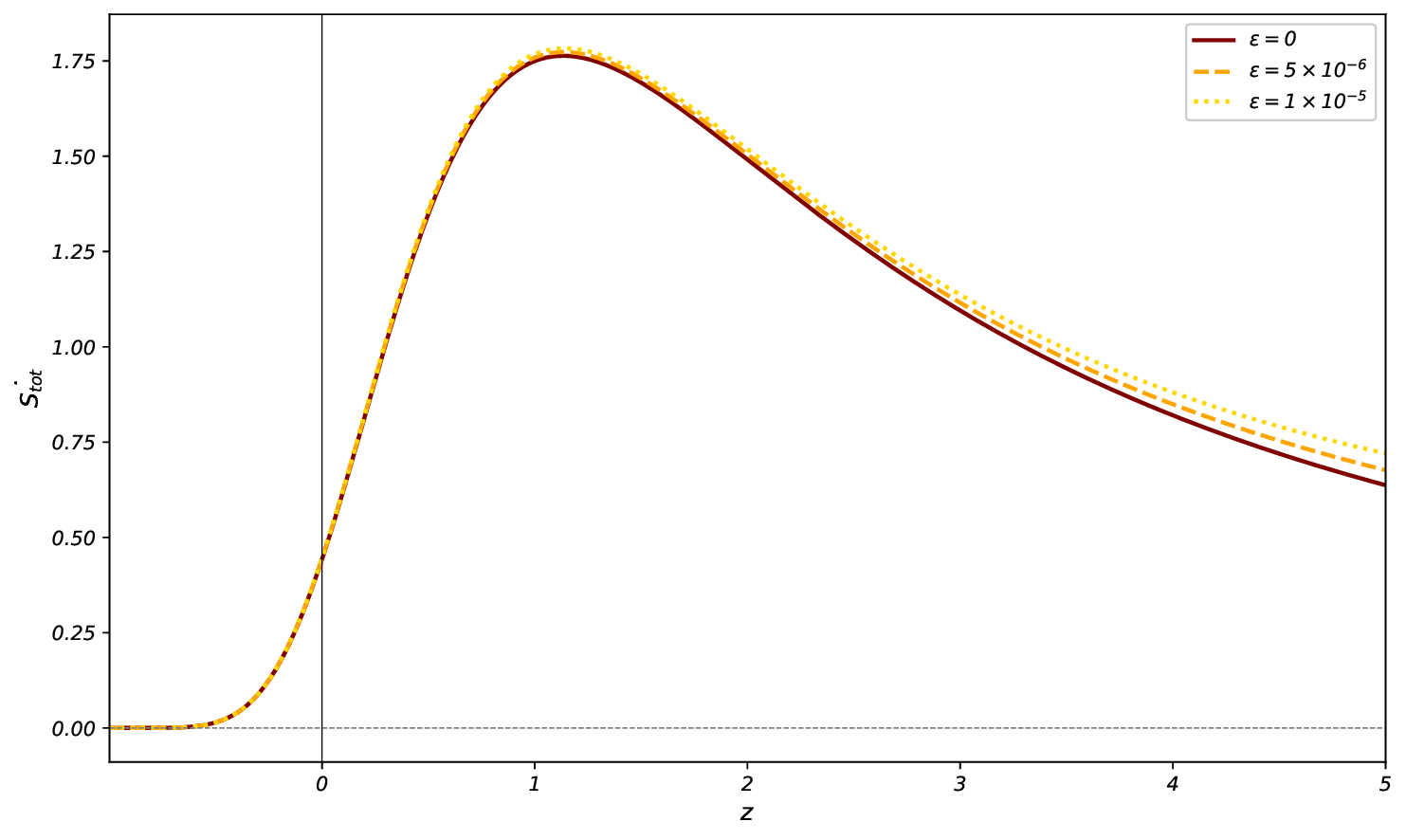}}
		\caption{Plot of the entropy production rate $\dot{S_\text{tot}}$ as a function of redshift $z$, including the Casimir energy contribution. The curves are obtained using the parameter constraints reported in Table~\ref{table2}. Three scenarios are shown: $\epsilon = 0$ corresponds to the $\Gamma$RVM model without the Casimir effect, while $\epsilon = 5\times10^{-6}$ and $\epsilon = 10^{-5}$ illustrate the impact of small but nonzero Casimir contributions. For larger values of $\epsilon$, the deviations become too large. The generalized second law (GSL) of thermodynamics is satisfied in all cases.}
    \label{fig20}
	\end{minipage}
\end{figure}

\noindent We want to check whether the generalized second law (GSL) of thermodynamics still holds when Casimir energy is taken into account. For that, we need to compute the rate of change of Casimir energy entropy $\dot{S_\mathrm{cs}}$. For this, we use the Gibbs equation given in Eq.~\eqref{Gibbs_eqn}:
\begin{equation}\label{gibbs_cs}
  T \dot{S_\mathrm{cs}} = (\rho_\mathrm{cs} + p_\mathrm{cs}) \dot{V} + \dot{\rho_\mathrm{cs}} V.
\end{equation}
Substituting the value for $\rho_\mathrm{cs}$ and the EoS parameter $w_\mathrm{cs} = 1/3$, we get
\begin{equation}\label{T_entropy_cs}
  T \dot{S_\mathrm{cs}} = \frac{16 \pi}{3} \rho_\mathrm{cs} [ {R_H}^2 \dot{R_H} - H {R_H}^3 ].
\end{equation}
Since we have assumed the temperature $T$ to be the Gibbs-Hawking temperature and the horizon radius is $1/H$, our expression for the rate of change of Casimir entropy can be simplified as:
\begin{equation}\label{entropy_cs}
  \dot{S_\mathrm{cs}} = - \frac{32 \pi^2}{3 H^3} \rho_\mathrm{cs} \left(1 + \frac{\dot{H}}{H^2} \right) = \frac{32 \pi^2}{3 H^3} \rho_\mathrm{cs} q ,
\end{equation}
where $q$ is the deceleration parameter. Since the total entropy of the Universe has a contribution from Casimir energy, the total change in entropy is now written as:
\begin{equation}\label{sdot_tot_cs}
  \dot{S_\text{tot}} = \dot{S_H} + \dot{S_m} + \dot{S_\mathrm{cs}}.
\end{equation}
We have already computed the horizon entropy $S_H$, along with the entropy of the dark fluid captured by $S_m$ in Sec~\ref{sec8.2}. For these components, the GSL was satisfied, evident from Eq.~\eqref{Sdot_total}. We now add the Casimir energy component calculated in Eq.~\eqref{entropy_cs} to the total entropy computed before to check if it still remains greater than zero. Therefore, we have:
\begin{equation}\label{Sdot_total_cs}
  \dot{S_\text{tot}} = \frac{8 \pi^2}{H}\left[2(1+q)+\frac{(1-\beta)\rho_m}{H^2}q + \frac{4 \rho_\mathrm{cs}}{3 H^2}q \right] .
\end{equation}
Substituting the value of $\rho_\mathrm{cs}$ from Eq.~\eqref{rho_cs},  we are left with a simplified expression as follows:
\begin{equation}\label{Sdot_total_cs_fin}
  \dot{S_\text{tot}} = \frac{8 \pi^2}{H} \left[2(1+q) - 2 q \frac{\dot{H}}{H^2} \right] =  \frac{16 \pi^2}{H}{(1+q)}^2 .
\end{equation}
We can see that this is the same expression we obtained in the case where the Casimir effect was not considered. Thus, the GSL of thermodynamics still holds, but due to the additional component of Casimir energy density, the Hubble parameter evolves differently, which affects the entropy evolution too. In particular, from Eq.~\eqref{fe_cs_diff} we obtain:
\begin{equation}\label{Hubble_cs}
\dot{H} = -\frac{\epsilon}{8\pi} (1+3\beta) H^{4} - \frac{3}{2}(1-\beta)(1-\nu)H^{2} + \frac{3}{2}\,(\Omega_{\Lambda}-\nu)(1-\beta)H_{0}^{2} .
\end{equation}
We remark on the $H^4$ term in the above equation, which is the minimum power of the Hubble Parameter required for inflation to occur and can be obtained by quantum field corrections in the RVM approach~\cite{Sola25}. Here, this term arises due to the Casimir effect, which is captured by the constant $\epsilon$. We note that the fourth power term vanishes when the Casimir energy is absent, i.e., when $\epsilon = 0$. Moreover, $\epsilon$ must admit very small values; of orders of magnitude less than $10^{-5}$, else the expansion blows up too rapidly. As remarked earlier, $\epsilon$ is a quantity tied to QFT and not explicitly derived. It has been found in laboratory experiments mimicking the Casimir effect in de Sitter space, that the Casimir energy term has a dependence on an IR-cutoff $L$ and a UV-cutoff $d$,~\cite{Li10} which are analogous to the apparent horizon radius $R_H$ and the Planck length $l_{\text{Pl}}$ respectively.\\
\indent Thus, the limit for $\epsilon$ remarked above is not a hard bound, and the value may change according to the choice of units. We have chosen natural units which allow $\epsilon$ to be a dimensionless constant. Our main concern was analyzing the GSL for the $\Gamma$RVM model with Casimir effect, which we have established still holds from Eq.~\eqref{Sdot_total_cs_fin}. We plot the evolution of $\dot{S_\text{tot}}$ considering the contribution of the Casimir effect with arbitrary values of $\epsilon$ in Fig.~\ref{fig20}. We fix the values of the model parameters from the constraints obtained for the BASE+CMB+$H_0$ dataset reported in Table~\ref{table2}. We can see from the trajectories that in the past, the entropy evolution diverges for varying Casimir effect influences. However, towards the future, all the evolutionary trajectories for $\dot{S_\text{tot}}$ combine as the value reaches zero. This implies that the total entropy converges to a maximum in the asymptotic limit $z \rightarrow -1$. These findings align with the general scenario and also suggest that the Casimir effect offers a promising avenue for investigating the $\Gamma$RVM framework.

\section{Conclusion}
\label{sec9}
In this work, we have studied cosmic expansion in a flat FLRW Universe with gravitationally induced adiabatic matter creation and a decaying vacuum, adopting $\rho_{\Lambda}=\tfrac{3}{8 \pi G}(c_0+\nu H^2)$ and $\Gamma=3 \beta H$. Using recent observational data, we have constrained the model parameters through MCMC analysis and examined the interplay within the dark sector. The results confirm an accelerated expansion phase and show consistency with $\Lambda$CDM across the three employed datasets.\\
\indent We have derived analytic expressions for the Hubble parameter $H(z)$, deceleration parameter $q(z)$, effective EoS $w_\text{eff}$, and transition redshift $z_\text{tr}$ for the $\Gamma$RVM model. We have tested the model against background observations from Pantheon SNe Ia, DESI BAO, and $H(z)$ data, and incorporated linear perturbation constraints from a $f(z)\sigma_8(z)$ compilation. We have further extended the analysis with an $H_0$ prior from SH0ES, and CMB distance priors from Planck 2018, noting that our parameters $\nu$ and $\beta$ are perturbative and recover $\Lambda$CDM in the zero limit.\\
\noindent One of the key results has been the confirmation of the Universe’s accelerated expansion. We have constrained the transition redshift $z_{\text{tr}}$, finding $z_{\text{tr}}=0.663 \pm 0.041$ (BASE), $z_{\text{tr}}=0.739^{+0.034}{-0.035}$ (BASE+$H_0$), and $z{\text{tr}}=0.697 \pm 0.015$ (BASE+CMB+$H_0$), with the $H_0$ prior shifting the transition to earlier times. We have also obtained the present effective EoS parameter $w^\text{eff}_0$, which lies between $-0.72$ and $-0.69$, consistent with late-time acceleration ($w\text{eff}<-1/3$) and indicating the ongoing transition from matter domination to vacuum domination as $w_\text{eff}\to -1$ in the future.\\
\indent For the fractional density parameter $\Omega_\Lambda$, we obtain mean values between $0.68$ and $0.72$, with higher estimates arising when the $H_0$ prior is included. While the first two datasets yield values slightly lower than $\Lambda$CDM expectations, including CMB data results in a marginally higher estimate within the $\Gamma$RVM framework. These results remain consistent with the standard cosmological model while supporting the case for a dynamical vacuum. We have also examined the $H_0$ tension: the estimates show discrepancies of $1.27\sigma$, $3.71\sigma$, and $1.45\sigma$ with respect to Planck 2018, and $3.22\sigma$, $2.02\sigma$, and 4.23$\sigma$ relative to the SH0ES value $H_0=73.04 \pm 1.04$ km/s/Mpc~\cite{Riess22}. The inclusion of the $H_0$ prior with the BASE dataset helps in easing the tension, though in the presence of CMB data, the improvement is less significant, with the residual tension only slightly reduced from 4.89$\sigma$.\\
\indent Another important aspect concerns the $S_8$ tension. To address this, we analyzed cosmic perturbations and structure formation through $\sigma_8$ and the derived parameter $S_8$. Figs.~\ref{fig14}–\ref{fig16} display the theoretical curves for the redshift evolution of the weighted linear growth rate function in both $\Lambda$CDM and $\Gamma$RVM, compared with 26 observational $f(z)\sigma_8(z)$ measurements. The curves show good agreement with the data. Our estimates of $S_8$ lie in the range $0.79$–$0.82$, consistent with the Planck value $S_8 = 0.831 \pm 0.013$, corresponding to a residual tension below $2\sigma$. While earlier weak lensing surveys suggested somewhat lower values ($S_8 < 0.78$), more recent surveys favor higher estimates, closer to the Planck result and in good concordance with our findings. \\
\indent We have further examined the evolution of the jerk parameter, which provides insight into whether the rate of the Universe’s acceleration is changing. In the standard $\Lambda$CDM scenario, the jerk parameter remains constant at $j(z)=1$ throughout cosmic history, reflecting a fixed rate of acceleration/deceleration. As expected for dynamical vacuum models, the $\Gamma$RVM slightly departs from this trajectory (Fig.\ref{fig17}). Eq.~\eqref{jerk_para_model} makes this deviation explicit. In particular, for the BASE+CMB+$H_0$ dataset, where $\beta$ and $\nu$ attain their smallest fitted values among the three data combinations, the departure from $j(z)=1$ is minimal. Moreover, the future evolution shows that the $\Gamma$RVM asymptotically approaches the standard behavior, with $j(z)\to 1$, thereby mimicking $\Lambda$CDM in the cosmic future.\\
\indent Our statistical analysis involved a direct comparison of the $\Gamma$RVM against the standard $\Lambda$CDM model. We first employed the reduced $\chi^2$ test and computed $\chi^2_{\rm red}$ for all three data combinations. For both models, the resulting values are close to $0.97$–$0.98$, indicating good agreement with the data but not allowing for a clear preference. To perform a more robust model selection, we further evaluated the Akaike Information Criterion (AIC) and the Deviance Information Criterion (DIC), as reported in Table~\ref{table3}. The model with the lower AIC (DIC) value is statistically favored. The differences, $\Delta$AIC and $\Delta$DIC, are defined relative to the $\Lambda$CDM baseline. According to $\Delta$AIC, the reference model is favored for the BASE dataset, whereas the inclusion of the $H_0$ prior and the subsequent augmentation with CMB data render the $\Gamma$RVM increasingly favorable. The $\Delta$DIC values for the first two datasets are nearly zero, implying no significant preference, but for the third dataset, the $\Gamma$RVM gains a slight advantage. Taken together, these results are encouraging and lend statistical support to the viability of the proposed model.\\
\indent Following the statistical analysis, we have carried out a thermodynamic study of the Universe within our proposed framework. Considering the modified matter–energy dynamics induced by particle creation, we employed the generalized first law of thermodynamics and derived expressions for the dark matter temperature and particle number density. To examine the entropy evolution, the Universe was modeled as a closed system and the generalized second law (GSL) of thermodynamics was analyzed. Under suitable assumptions, we obtained analytic expressions for the total entropy $S$ and its time derivative $\dot{S}$. Using the constraints obtained from the MCMC analysis, we verified that the proposed model respects the GSL and the convexity condition, ensuring thermodynamic stability. Furthermore, we extended the analysis by incorporating the Casimir effect into the proposed cosmological scenario. The GSL remains valid in this case, highlighting the framework's robustness and suggesting an intriguing direction for future investigations.\\
\indent In conclusion, the phenomenological model depicting the interaction between dark matter with matter creation and decaying vacuum dark energy, as explored in the paper, effectively describes the evolutionary features of the Universe at both background and perturbation levels. The thermodynamic features are also in accordance with established laws. Thus, exploring an interacting model in which dark matter undergoes matter creation and couples to a running vacuum term draws on multiple perspectives, including particle production by the gravitational field and quantum vacuum fluctuations. Inclusion of the Casimir effect provides a coherent picture. These avenues have been rigorously investigated and continue to be active research areas. Overall, our results underscore the viability of the $\Gamma$RVM framework as a compelling alternative to $\Lambda$CDM, capable of addressing fundamental cosmological puzzles while maintaining agreement with current observational data. \\\\

\acknowledgments
L.C. would like to thank University Grant Commission (UGC), India for providing Junior Research Fellowship (JRF) to carry out this work.

%\paragraph{Note added.} This is also a good position for notes added
%after the paper has been written.

\end{document}